\documentclass[longauth]{aa} 

\usepackage{graphicx}
\usepackage{txfonts}
\usepackage{natbib}
\usepackage[colorlinks=true,linkcolor=bluea,allcolors=blue]{hyperref}
\begin{document}

   \title{A biconical ionised gas outflow and evidence for positive feedback in NGC 7172 uncovered by MIRI/JWST}
   \titlerunning{A biconical outflow and evidence for positive feedback in NGC 7172 uncovered by MIRI/JWST}

   \author{L. Hermosa Mu{\~n}oz \inst{1}
        \and
        A. Alonso-Herrero \inst{1}
        \and
        M. Pereira-Santaella\inst{2}
        \and
        I. Garc{\'i}a-Bernete\inst{3}
        \and
        S. Garc{\'i}a-Burillo\inst{4}
        \and
        B. Garc{\'i}a-Lorenzo\inst{5,6}
        \and 
        R. Davies\inst{7}
        \and 
        T. Shimizu\inst{7}
        \and 
        D. Esparza-Arredondo\inst{5,6}
        \and
        E. K. S. Hicks\inst{8,9,10}
        \and
        H. Haidar\inst{11}
        \and
        M. Leist\inst{9}
        \and
        E. L{\'o}pez-Rodr{\'i}guez\inst{12}
        \and
        C. Ramos Almeida\inst{5,6}
        \and
        D. Rosario\inst{11}
        \and
        L. Zhang\inst{9}
        \and
        A. Audibert\inst{5,6}
        \and
        E. Bellocchi\inst{13,14}
        \and
        P. Boorman\inst{15}
        \and 
        A. J. Bunker\inst{3}
        \and
        F. Combes\inst{16}
        \and
        S. Campbell\inst{11}
        \and
        T. D{\'i}az-Santos\inst{17,18}
        \and
        L. Fuller\inst{9}
        \and
        P. Gandhi\inst{19}
        \and
        O. Gonz{\'a}lez-Mart{\'i}n\inst{20}
        \and
        S. Hönig\inst{19}
        \and
        M. Imanishi\inst{21,22}
        \and
        T. Izumi\inst{22,22}
        \and
        A. Labiano\inst{23}
        \and
        N.A. Levenson\inst{24}
        \and
        C. Packham\inst{9}
        \and
        C. Ricci\inst{25,26}
        \and
        D. Rigopoulou\inst{3,18}
        \and
        D. Rouan\inst{27}
        \and 
        M. Stalevski\inst{28,29}
        \and
        M. Villar-Mart{\'i}n\inst{30}
        \and
        M. J. Ward\inst{31}
        }

   \institute{
        1. Centro de Astrobiolog{\'i}a (CAB) CSIC-INTA, Camino bajo el Castillo s/n, 28692, Villanueva de la Cañada, Madrid, Spain \\
              \email{lhermosa@cab.inta-csic.es}
         \\
        2. Instituto de F{\'i}sica Fundamental, CSIC, Calle Serrano 123, E-28006, Madrid, Spain
             \\
        3. Department of Physics, University of Oxford, Keble Road, Oxford, OX1 3RH, UK \\
        4. Observatorio Astron{\'o}mico Nacional (OAN-IGN) - Observatorio de Madrid, Alfonso XII, 3, 28014, Madrid, Spain \\
        5. Instituto de Astrof{\'i}sica de Canarias, C/ V{\'i}a L{\'a}ctea s/n, 38205 La Laguna, Tenerife, Spain \\
        6. Departamento de Astrof{\'i}sica, Universidad de La Laguna, 38205 La Laguna, Tenerife, Spain \\
        7. Max Planck Institute for Extraterrestrial Physics (MPE), Giessenbachstr.1, 85748 Garching, Germany \\
        8. Department of Physics and Astronomy, University of Alaska Anchorage, Anchorage, AK 99508-4664, USA \\
        9. Department of Physics and Astronomy, The University of Texas at San Antonio, 1 UTSA Circle, San Antonio, Texas, 78249, USA \\
        10. Department of Physics, University of Alaska, Fairbanks, Alaska 99775-5920, USA \\
        11. School of Mathematics, Statistics and Physics, Newcastle University, Newcastle upon Tyne, NE1 7RU, UK \\
        12. Kavli Institute for Particle Astrophysics \& Cosmology (KIPAC), Stanford University, Stanford, CA 94305, USA \\
        13. Departmento de Física de la Tierra y Astrof{\'i}sica, Fac. de CC F{\'i}sicas, Universidad Complutense de Madrid, E-28040 Madrid, Spain \\
        14. Instituto de F{\'i}sica de Part{\'i}culas y del Cosmos IPARCOS, Fac. CC F{\'i}sicas, Universidad Complutense de Madrid, E-28040 Madrid, Spain \\
        15. Cahill Center for Astrophysics, California Institute of Technology, 1216 East California Boulevard, Pasadena, CA 91125, USA \\
        16. LERMA, Observatoire de Paris, Collège de France, PSL University, CNRS, Sorbonne University, France, Paris \\
        17. Institute of Astrophysics, Foundation for Research and Technology-Hellas (FORTH), Heraklion 70013, Greece \\ 
        18. School of Sciences, European University Cyprus, Diogenes street, Engomi, 1516 Nicosia, Cyprus \\ 
        19. School of Physics \& Astronomy, University of Southampton, Highfield, Southampton SO17 1BJ, UK \\ 
        20. Instituto de Radioastronom{\'i}a y Astrof{\'i}sica (IRyA), Universidad Nacional Aut{\'o}noma de M{\'e}xico, Antigua Carretera a P{\'a}tzcuaro 8701, ExHda. San Jos{\'e} de la Huerta, Morelia, Michoac{\'a}n, M{\'e}xico C.P. 58089 \\
        21. National Astronomical Observatory of Japan, National Institutes of Natural Sciences (NINS), 2-21-1 Osawa, Mitaka, Tokyo 181-8588, Japan \\ 
        22. Department of Astronomy, School of Science, Graduate University for Advanced Studies (SOKENDAI), Mitaka, Tokyo 181-8588, Japan \\ 
        23. Telespazio UK for the European Space Agency (ESA), ESAC, Camino Bajo del Castillo s/n, 28692 Villanueva de la Ca{\~n}ada, Spain \\
        24. Space Telescope Science Institute, 3700 San Martin Drive, Baltimore, MD 21218, USA \\ 
        25. Instituto de Estudios Astrof{\'i}sicos, Facultad de Ingenier{\'i}a y Ciencias, Universidad Diego Portales, Av. Ej{\'e}rcito Libertador 441, Santiago, Chile \\
        26. Kavli Institute for Astronomy and Astrophysics, Peking University, Beijing 100871, China \\ 
        27. LESIA, Observatoire de Paris, Université PSL, CNRS, Sorbonne Université, Sorbonne Paris Citeé, 5 place Jules Janssen, F-92195 Meudon, France \\ 
        28. Astronomical Observatory, Volgina 7, 11060 Belgrade, Serbia \\
        29. Sterrenkundig Observatorium, Universiteit Gent, Krijgslaan 281-S9, Gent B-9000, Belgium \\
        30. Centro de Astrobiolog{\'i}a (CAB), CSIC-INTA, Ctra. de Ajalvir, km 4, 28850 Torrej{\'o}n de Ardoz, Madrid, Spain \\
        31. Centre for Extragalactic Astronomy, Durham University, South Road, Durham DH1 3LE, UK
        } 

   \date{Received M DD, YYYY; accepted M DD, YYYY}

  \abstract{The type-2 Seyfert NGC\,7172 hosts one of the lowest ionised gas mass outflow rates ($\dot{M}_{out} \sim$\,0.005\,M$_{\sun}$\,yr$^{-1}$) in a sample of six active galactic nuclei (AGN) with similar bolometric luminosities (log L$_{bol} \sim 44$\,erg\,s$^{-1}$) within the Galactic Activity, Torus and Outflow Survey (GATOS).}{We present observations of NGC\,7172 obtained with the medium-resolution spectrometer (MRS) of the Mid-Infrared Instrument (MIRI) on board of the James Webb Space Telescope (JWST). We aim to understand the properties of the ionised gas outflow and its impact on the host galaxy.}{We mainly used the ionised gas emission lines from the neon transitions, that cover a broad range of ionisation potentials (IP) from $\sim$20\,eV to $\sim$130\,eV. We applied parametric and non-parametric methods to characterise the line emission and kinematics.}{The low excitation lines (IP\,$<$\,25\,eV, e.g. [Ne\,II]) trace the rotating disc emission. The high excitation lines (IP\,$>$\,90\,eV, e.g. [Ne\,V]), which are likely photoionised exclusively by the AGN, are expanding in the direction nearly perpendicular to the disc of the galaxy, with maximum projected velocities in the range of $\sim$350-500\,km\,s$^{-1}$. In particular, [Ne\,V] and [Ne\,VI] lines reveal a biconical ionised gas outflow emerging north-south from the nuclear region, extending at least $\sim$2.5\arcsec\,N and 3.8\arcsec\,S (projected distance of $\sim$\,450 and 680\,pc, respectively). Most of the emission arising in the northern part of the cone was not previously detected due to obscuration.}{Given the almost face-on orientation of the outflow and the almost edge-on orientation of the galaxy, NGC\,7172 may be a case of weak coupling. Nevertheless, we found evidence for positive feedback in two distinct outflowing clumps at projected distances of 3.1\arcsec\,and 4.3\arcsec\,(i.e. $\sim$560 and 780\,pc) south-west from the AGN. We estimated a star formation rate in these regions using the [Ne\,II] and [Ne\,III] luminosities of 0.08\,M$_{\sun}$\,yr$^{-1}$, that is $\sim$10\% of that found in the circumnuclear ring. The star formation activity might have been triggered by the interaction between the ionised gas outflow and the interstellar medium of the galaxy.} 


   \maketitle
%

\section{Introduction}
\label{Sect1:Introduction}

The evolution of galaxies during cosmic time is believed to be impacted up to a certain degree by feedback processes, particularly for those harbouring an active galactic nucleus (AGN) \citep[see e.g.][]{Kormendy2013,Harrison2018,Harrison2024}. These processes may play a crucial role in altering the gas content of galaxies through driving it outwards in the so-called outflows, or by the accretion of new gas from the intergalactic medium, the so-called inflows \citep[see][]{Davies2014,StorchiBergmann2019}. Several works in the literature have explored the important question of the impact of outflows on the ability to form new stars in their host galaxies \citep[see e.g.][]{Cresci2015a,Harrison2017,Venturi2021,Veilleux2020,Speranza2024}. On the one hand, if the gas is heated or expelled from the galaxy as outflows, these could potentially prevent star formation \citep[i.e. negative feedback; e.g.][]{Harrison2017}. On the other hand, if the gas cools down due to fragmentation of the outflow or compression of molecular clouds, star formation could be triggered, even out of the galaxy plane \citep[i.e. positive feedback; e.g.][]{Silk2013,Cresci2015b,Gallagher2019,Bellocchi2023}. There are reported cases of positive feedback occurring at low \citep{Maiolino2017,Shin2019,Perna2020,Venturi2023} and high redshifts \citep[e.g.][]{Cresci2015a,Nesvadba2020}, although it is a relatively new observational phenomena. Thus, the complete picture of the impact of AGN-launched outflows into their host galaxies is still not clear, despite several attempts in the literature \citep{Fiore2017,Davies2020}.

NGC\,7172 is an almost edge-on spiral galaxy with a type-2 Seyfert nucleus \citep{Veron2006}, log L$_{bol} \sim 44.1$\,erg\,s$^{-1}$ \citep{Davies2020}, which has been widely studied in the literature \citep[see e.g.][]{Sharples1984,Thean2000,Smajic2012,Thomas2017,AH2023}. It is a nearby object located at a redshift of 0.00868 (\href{https://ned.ipac.caltech.edu/}{NASA/IPAC Extragalactic Database; NED}) and a redshift independent distance of 37\,Mpc \citep[1\arcsec$\sim$180\,pc, see][]{Davies2015}. The galaxy hosts a circumnuclear ring, with a diameter of approximately 1.1-1.4\,kpc as observed with the CO(3$-$2) transition \citep[345\,GHz;][from now on AH23]{AH2023}. A dust lane crossing the major axis covers emission from the central and northern regions, especially in the optical range. \cite{Thomas2017} detected an ionisation cone in this range with an opening angle of 120$^{\circ}$ which appears to be extending out of the galaxy mostly detected to the south in the optical, with the northern part nearly fully obscured by the host galaxy. \cite{Davies2020} derived the nuclear properties of the ionised gas outflow using X-shooter optical data, and obtained a relatively low ionised mass outflow rate, $\dot{M}_{\rm out} \sim$\,0.005\,M$_{\odot}$\,yr$^{-1}$. \cite{AH2023} also detected the ionised gas outflow in the near-infrared (near-IR) using the [Si\,VI]\,1.96$\mu$m line, and concluded that it is causing the molecular gas ring to outflow in the plane of the galaxy \citep[see also][]{Stone2016}. There are however no detections of an outflow in X-rays \citep{Igo2020}.

We observed NGC\,7172 with the mid-infrared spectrometer (MRS) of the Mid-Infrared Instrument \citep[MIRI;][]{Wells2015,Rieke2015,Wright2015,Wright2023} on board of the James Webb Space Telescope \citep[JWST;][]{Gardner2023}. This data set was observed within the Galactic Activity, Torus and Outflow Survey (\href{https://gatos.myportfolio.com/}{GATOS}) collaboration \citep{GarciaBurillo2021,AH2021}, with the aim to understand the relation of the outflow properties in a sample of six hard X-ray selected nearby AGN with similar luminosities \citep[log\,$L_{AGN} \sim 44$\,erg\,s$^{-1}$,][Zhang et al. submitted.]{Davies2020}. Among these galaxies, NGC\,7172 shows one of the lowest ionised mass outflow rates \citep[see][]{Davies2020}. In this paper we present the analysis of the MIRI data on NGC\,7172 (Sect.~\ref{Sect2:Data}) showing the main results (i.e. kinematic, flux, and line ratios maps; Sect.~\ref{Sect3:Results}). We also discuss the detection of an ionised gas outflow and the possible existence of positive AGN feedback (Sects.~\ref{Sect4:Discussion} and~\ref{Sect5:Conclusions}). 

\section{Data and methodology}
\label{Sect2:Data}

The target was observed with the medium-resolution spectrometer (MRS) of MIRI/JWST within the program ID 1670 (P.I.:~T.~Shimizu and~R.~Davies). For the data reduction, we primarily followed the standard MRS pipeline procedure (e.g., \citealt{Labiano2016}; \citealt{Bushouse2023} and references therein) and the same configuration of the pipeline stages described in \citet{GarciaBernete2022c} and \citet{Pereira-Santaella2022} to reduce the data using the pipeline release 1.11.4 and the calibration context 1130. Some hot and cold pixels are not identified by the current pipeline version, so we added an extra step before creating the data cubes to mask them. The data reduction and extra steps are described in detail in \citet{GarciaBernete2024}, where the nuclear spectrum of NGC\,7172 can be seen in their Fig.~1 \citep[see also][]{Pereira-Santaella2022,GarciaBernete2022c}.
The total wavelength range of the MRS goes from 4.9$\mu$m to 28$\mu$m, divided in four channels with different resolutions and field-of-views (FoV; see details in \citealt{Pereira-Santaella2022,GarciaBernete2024}).

In this work we individually modelled the main ionisation emission lines in the data cubes using the \textsc{lmfit} package (v1.2.2) inside a \textsc{Python} environment (v3.9.12). We fitted each emission line using a combination of a linear fit of the continuum around each line ($\sim$\,0.03\,$\mu$m for each side) with at least one Gaussian component and a maximum of three per line. To select the adequate number of Gaussians that are sufficient to obtain a proper modelling to the lines, we used the $\varepsilon$ criteria described in \cite{Cazzoli2018} \citep[see also][]{HM2024}. In short, if the residuals of the continuum in the line after the Gaussian modelling are larger than three times the residuals of the continuum out of the line, then an extra component is added. When using more than one Gaussian, they are ordered to ensure continuity in the kinematic and flux maps, typically assuming the primary component to be the narrowest. We imposed the signal-to-noise (S/N) of the peak of the lines to be $>$\,3. After the modelling, we corrected the velocity dispersion from its instrumental value \citep{Argyriou2023} using $\sigma = \sqrt{\sigma^{2}_{\rm obs} - \sigma^{2}_{\rm inst}}$. The maps with the modelling with a single Gaussian for the neon emission lines are shown in Fig.~\ref{Fig:KinMaps_1comp}, and their corresponding continuum maps are in Fig.~\ref{FigAp:ContinuumMaps}.

Besides the parametric modelling, we applied a non-parametric modelling to the emission lines following a method similar to \cite{Harrison2014}. Contrary to this work, we do not use a previous Gaussian modelling to define the line, but rather we define the line as where the flux is higher than three times the standard deviation of the continuum near the line (from $\pm$700 to $\pm$1000\,km\,s$^{-1}$, see example in Fig.~\ref{FigAp:FitExamples}). Then we measure the parameters v02, v10, v50, v90, and v98, defined as the velocity corresponding to the 2\%, 10\%, 50\%, 90\%, and 98\% of the total accumulated flux from the line, without assuming any prior conditions to characterise the profiles (i.e. parameters from Gaussian profiles). The v02 and v98 parameters are indicators of the maximum and minimum velocities of the line, and thus are typically used in the literature as measurements of the outflow velocities \citep[see e.g.][]{Speranza2024}. In general, the results from the parametric and non-parametric modellings are consistent \citep[see][for a detailed comparison between both methods]{Hervella2023}. We used the parametric modelling for the spaxel-by-spaxel analysis, and the non-parametric method only for modelling the profiles from integrated spectra in some regions of interest. 

Given that the kinematic results for the lines are consistent for different ionisation potentials (IP; see Sect.~\ref{Subsec3:Results_nonparam}), in this work we will focus mostly on a limited number of them tracing gas emission from low to high ionisation states. We selected the neon emission lines, particularly [Ne\,II]\,12.81\,$\mu$m, [Ne\,III]\,15.55\,$\mu$m, [Ne\,V]\,14.32\,$\mu$m (hereafter [Ne\,V] unless specified otherwise)\footnote{We also detected the [Ne\,V]24.32$\mu$m, but we will focus on the [Ne\,V]14.32$\mu$m as it has better spatial and spectral resolution.}, and [Ne\,VI]\,7.65\,$\mu$m, as they all are bright lines from the same element with different IPs (see Table~\ref{Table:fluxes}) and thus and we can exploit their line ratios to further understand the physical origin of the gas (estimation of densities, etc.). The kinematic maps of the mentioned neon lines obtained with a parametric modelling with a single Gaussian component are shown in Fig.~\ref{Fig:KinMaps_1comp}. 

We also generated spatially-resolved line-ratios with the neon emission lines flux maps (i.e. [Ne\,II], [Ne\,III], and [Ne\,V]) fitted with one Gaussian (see Sect.~\ref{SubSect3:Results_Morph}). These lines are all located in channel 3 (FoV 5.2\arcsec$\times$6.2\arcsec), thus the proximity in wavelength of these lines ensures that the ratios are not significantly dependent on the extinction \citep{Tommasin2008}.

\section{Results}
\label{Sect3:Results}

\begin{figure*}
 \centering
    \includegraphics[width=.9\textwidth]{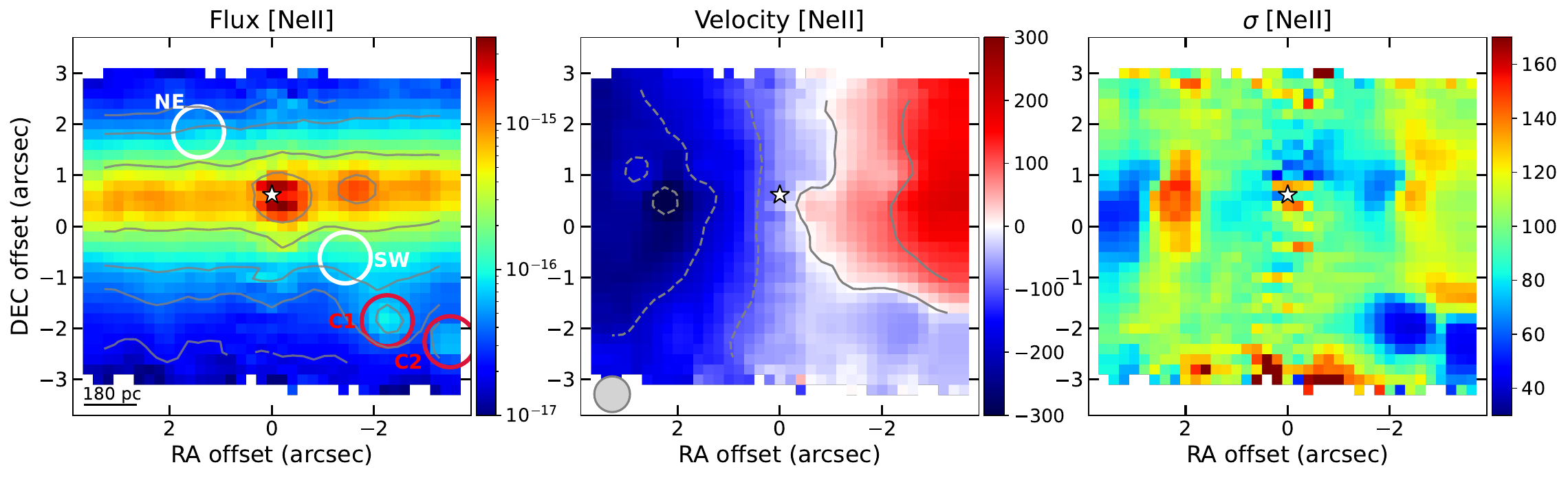}
    \includegraphics[width=.9\textwidth]{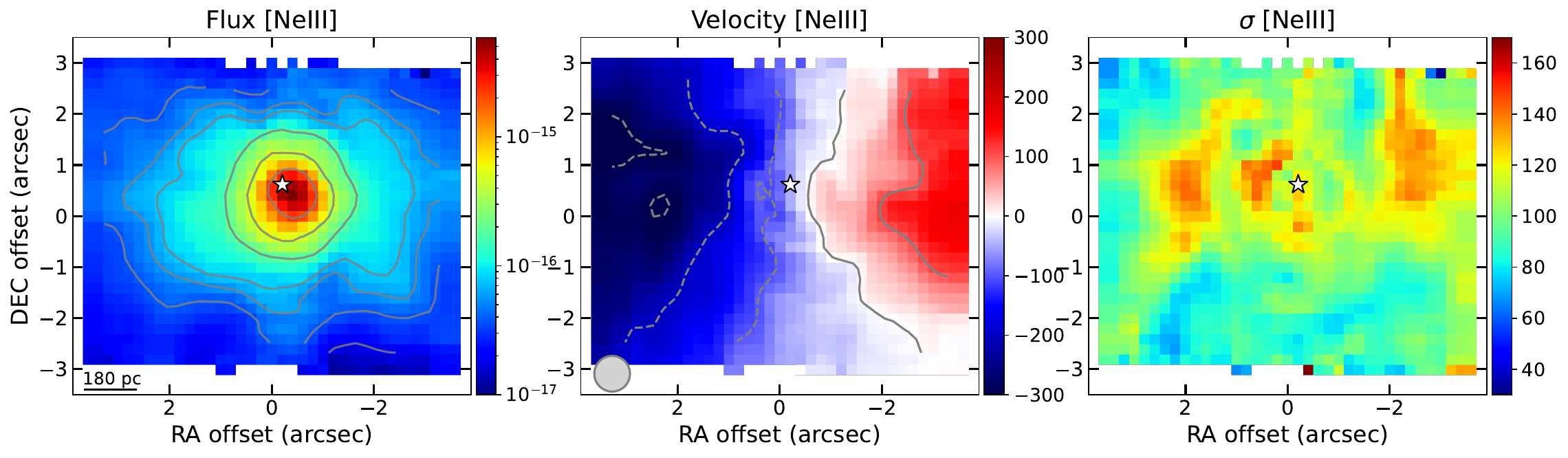}
    \includegraphics[width=.9\textwidth]{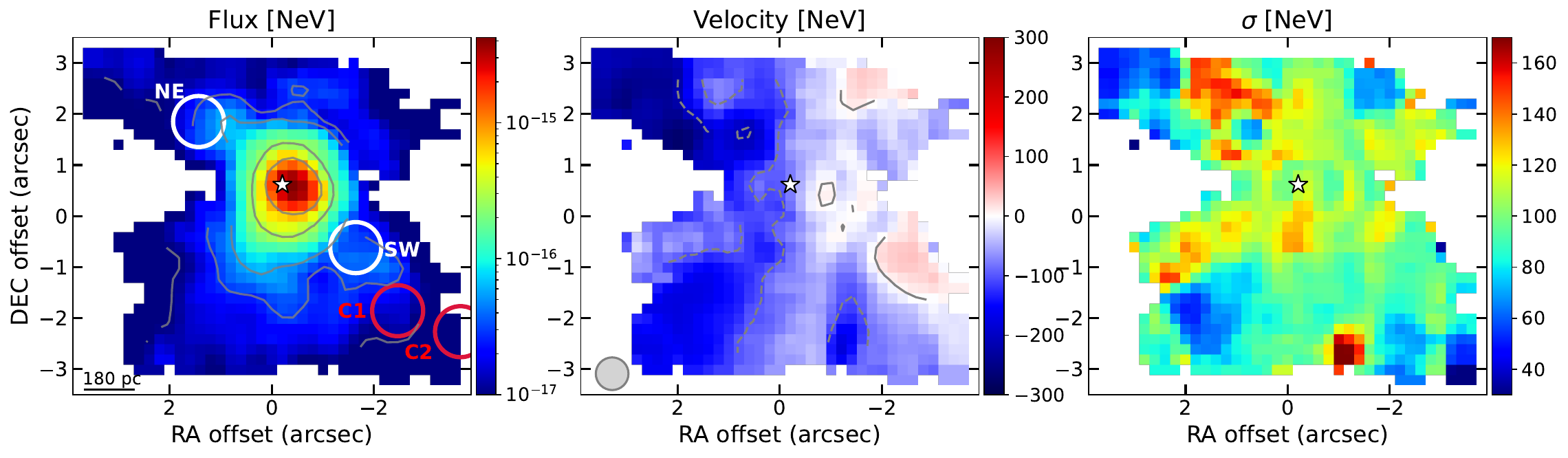}
    \includegraphics[width=.9\textwidth]{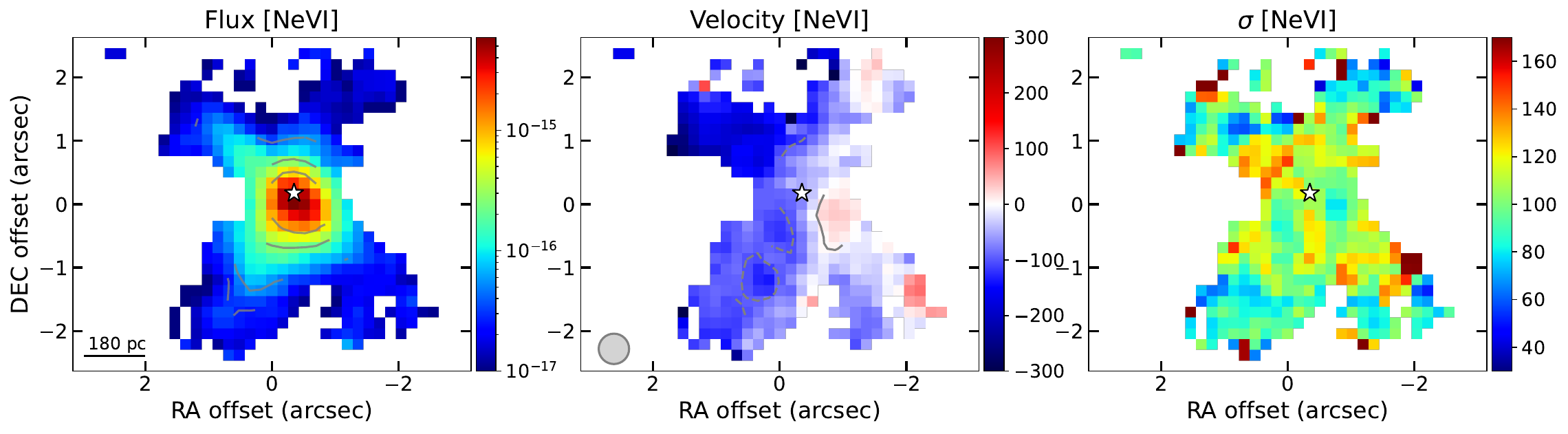}
     \caption{Maps of (from top to bottom) the [Ne\,II], [Ne\,III], [Ne\,V] and [Ne\,VI] emission lines fitted with one Gaussian component (S/N\,$>$\,3 at the line peak, see Sect.~\ref{Sect2:Data}). From left to right we show the flux maps in erg\,s$^{-1}$\,cm$^{-2}$, the mean velocity, and the velocity dispersion, $\sigma$ (corrected from the instrumental value), in km\,s$^{-1}$. The photometric centre of the continuum for each channel is marked with a white star. The 1\arcsec\,physical scale is marked with a black line in the left panels. The white and red circles in the [Ne\,II] and [Ne\,V] flux maps indicate the integrated regions discussed in the text (see Sect.~\ref{Sect3:Results}). The FWHM of the PSF is shown as a grey circle in the middle panels. The contours in the velocity maps go from -300 to 300\,km\,s$^{-1}$. In all panels, north (east) is up (left).}
    \label{Fig:KinMaps_1comp}
 \end{figure*}

The majority of the emission lines show signatures of two different kinematic components (see examples in Figs.~\ref{Fig:IntProfiles} and~\ref{FigAp:IntProfiles_perchannel}). In the central part of the galaxy, the line profiles are symmetric and can be modelled adequately with a single Gaussian. However, further out we detected wings in the lines both to the blue and to the red (see left and right panels of Fig.~\ref{Fig:IntProfiles}). To illustrate this, we selected two regions away from the diffraction ring of the point-spread function (PSF, see Fig.~\ref{FigAp:ContinuumMaps}) at both sides of the AGN towards the blueshifted and redshifted parts of the velocity field for all the emission lines. They are located at $\sim$3\arcsec\,north-east (NE; i.e. projected distance $\sim$540\,pc) and $\sim$2\arcsec\,south-west (SW; i.e. projected distance $\sim$360\,pc) from the centre (white circles in Fig.~\ref{Fig:KinMaps_1comp}). 

\subsection{Non-parametric modelling}
\label{Subsec3:Results_nonparam}

\begin{figure*}
 \centering
    \includegraphics[width=.675\columnwidth]{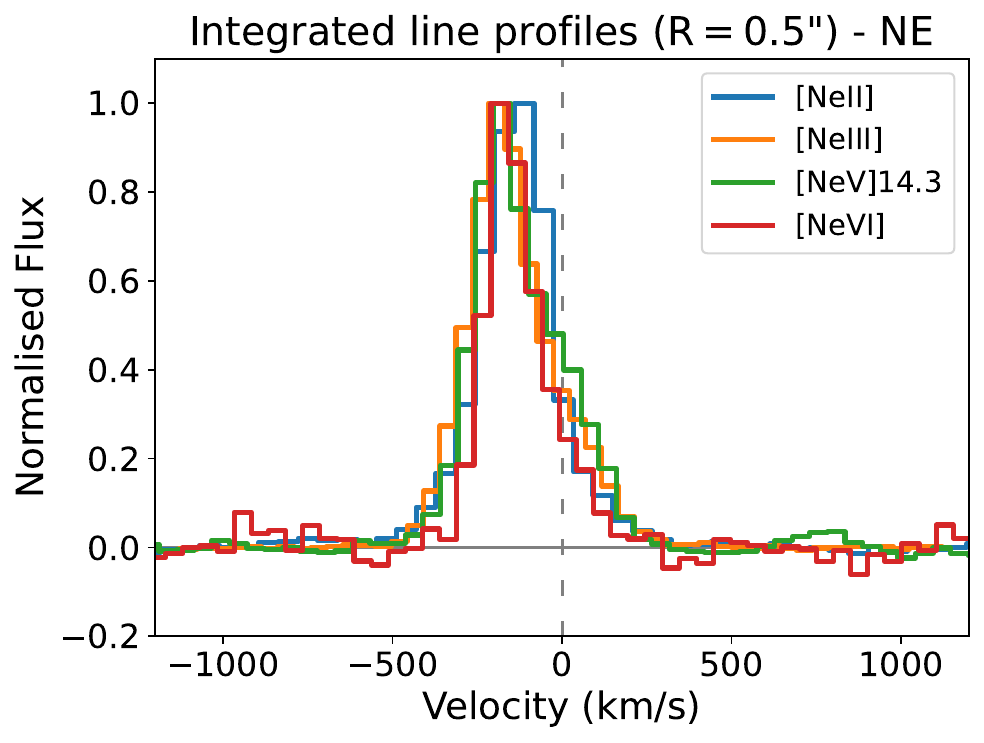}
    \includegraphics[width=.675\columnwidth]{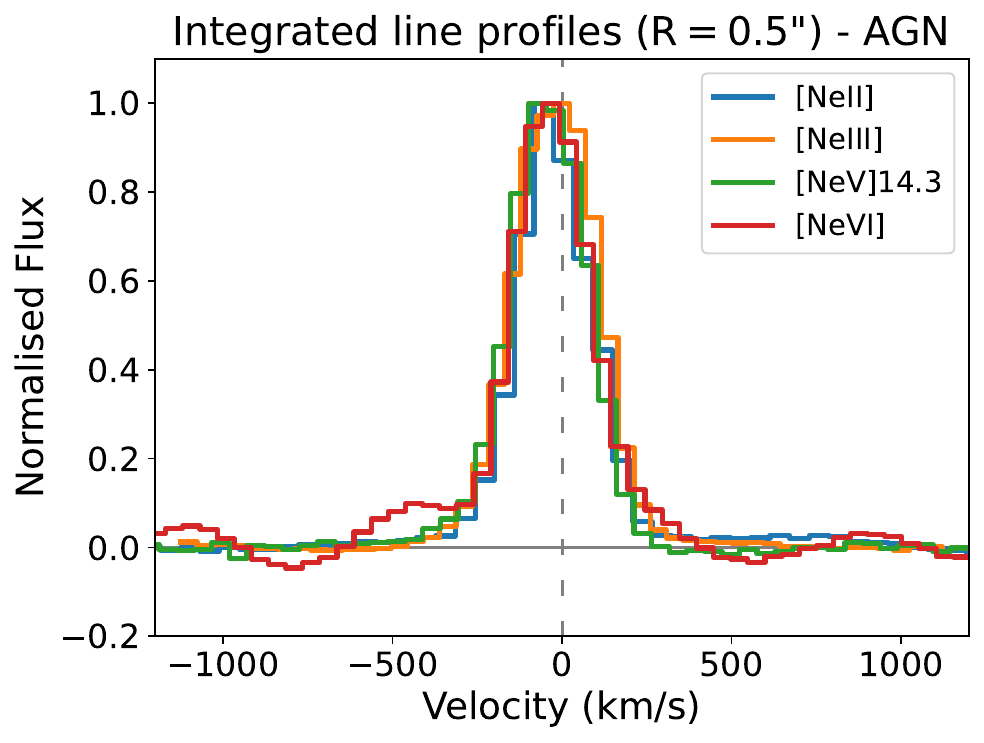}
    \includegraphics[width=.675\columnwidth]{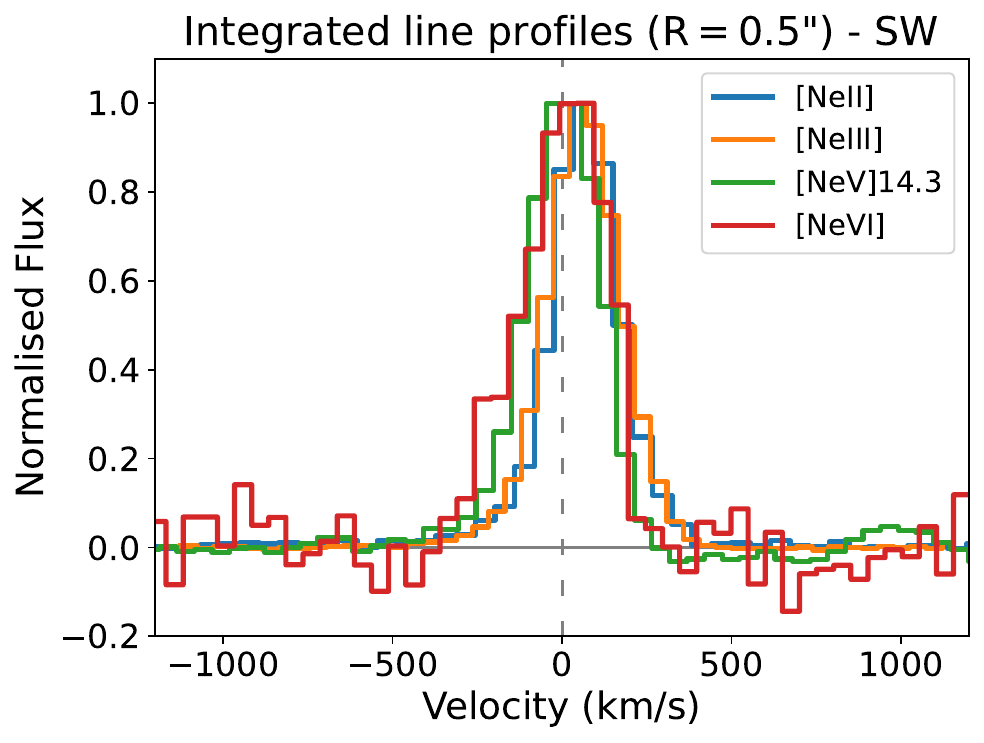}
     \caption{Emission line profiles of the ionised gas obtained from an integrated spectrum (circular aperture with R$\sim$0.5\arcsec) in the nuclear region (middle), NE (left) and SW (right) parts of the cone (see Fig.~\ref{Fig:KinMaps_1comp}). All the velocities refer to the systemic value computed with z\,$=$\,0.00868 (see Sect.~\ref{Sect1:Introduction}) indicated by the dashed, vertical, grey lines. Fluxes were normalised to the maximum of each line.}
    \label{Fig:IntProfiles}
 \end{figure*}

Previous works in the literature have reported a correlation between the width of the line with the IP \citep[see e.g.][]{Dasyra2011}, or the critical density \citep[see e.g.][]{DeRobertis1986}. 
Using the non-parametric method (see Sect.~\ref{Sect2:Data}), we measured the widths of all the ionised emission lines detected in the integrated spectra to check if such a trend is also present for NGC\,7172. We estimated the width of the nuclear integrated line profiles (see Figs.~\ref{Fig:IntProfiles} and~\ref{FigAp:IntProfiles_perchannel}) with the W80 parameter \citep{Harrison2014}, defined as $|v10-v90|$, and equal to $1.088$ times the full-width at half maximum (FWHM) for a Gaussian profile\footnote{The W80 parameter is not corrected by the instrumental resolution.}. We estimated the errors accounting for the standard deviation of the continuum near the line (width of $\sim$100\,km\,s$^{-1}$ per each side, from $\pm 800$\,km\,s$^{-1}$) and the spectral resolution per each line \citep{Argyriou2023}. 
Additionally, for the emission lines in the integrated nuclear spectrum, we performed a Monte-Carlo test with 30000 iterations to further constrain the uncertainties from the W80 measurements coming from the continuum noise (see more details in Appendix~\ref{Appendix1}). The W80 measurements obtained with this technique are consistent with the reported values in Table~\ref{Table:W80} in all cases. For the lines with the highest S/N, as [Ne\,II], [Ne\,III], or [Ne\,V]$14\mu$m, the statistical uncertainties are smaller than 10\,km\,s$^{-1}$, and up to $\sim 50$\,km\,s$^{-1}$ for lines with the lowest S/N, such as [Fe\,VIII] (S/N\,$\sim 5$, see upper, middle panel in Fig.~\ref{FigAp:IntProfiles_perchannel}). 

\begin{figure}
    \includegraphics[width=\columnwidth]{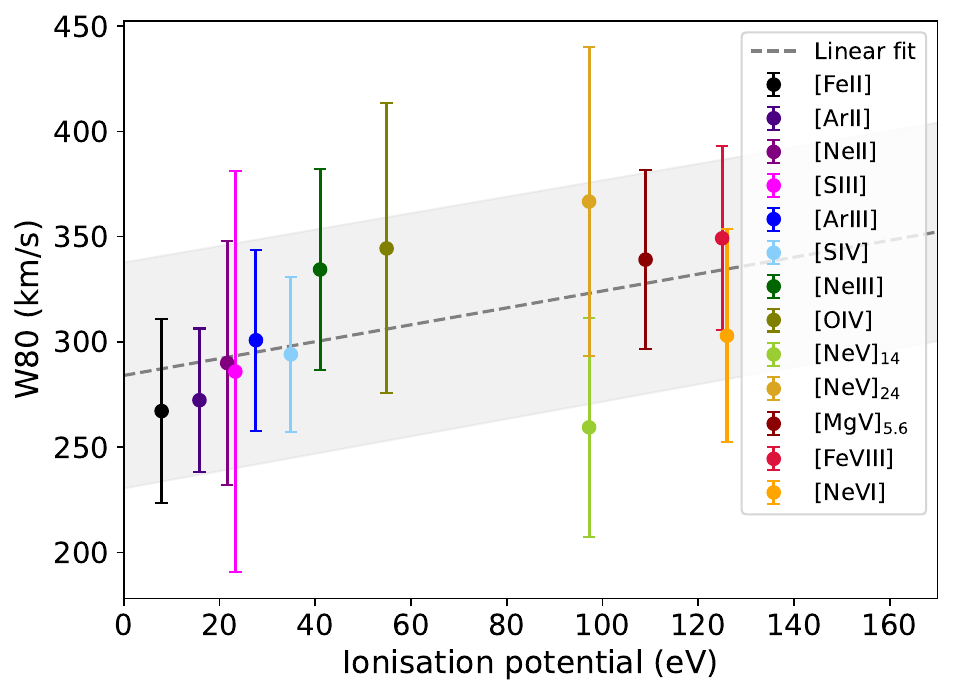}
    \includegraphics[width=\columnwidth]{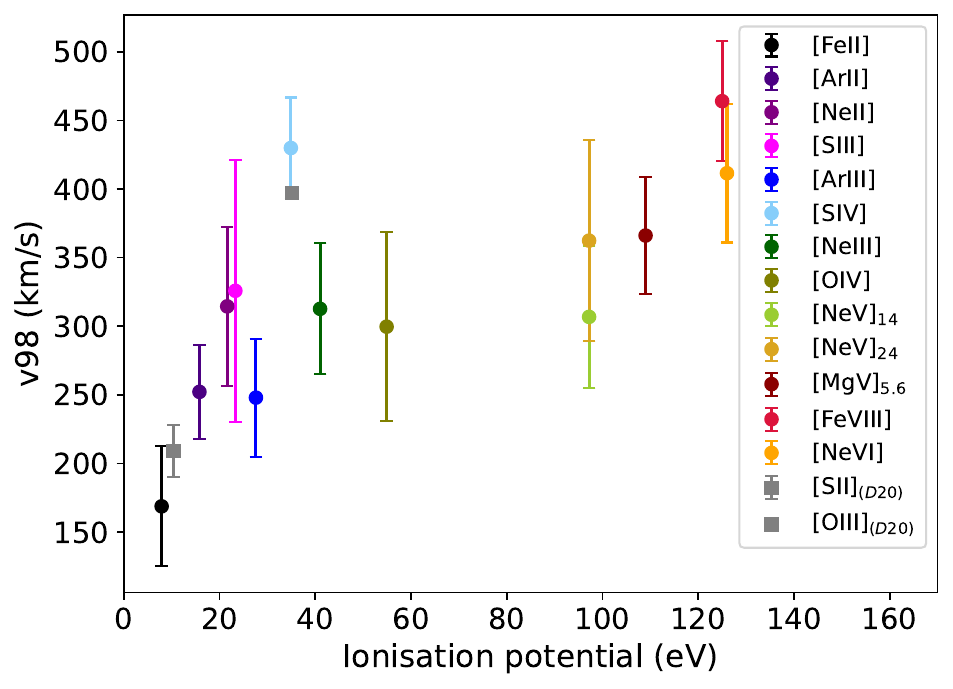}
     \caption{W80 (top panel) and v98 (bottom panel) measured in the integrated nuclear spectrum (R$\sim$0.5\arcsec) with respect to the emission lines IP. The dashed, grey line in the top panel corresponds to a linear fit of the points and the 1$\sigma$ uncertainty. The grey squares  in the bottom panel indicate the v98 value reported by \cite{Davies2020} for the optical $[$S\,II$]$ and $[$O\,III$]$ lines with their uncertainties. The errorbars in our data are obtained accounting for the standard deviation of the continuum near the line and the spectral resolution.}
    \label{Fig:W80V98}
 \end{figure}

The results, represented in Fig.~\ref{Fig:W80V98}, mainly show that all the emission lines with similar IPs behave similarly. There is a slight trend of increasing W80 at higher IPs (slope 0.4$\pm$0.2), although the correlation is moderate (Pearson coefficient $\rho \sim 0.5$ and p-value\,$\sim 0.08$, see Table~\ref{Table:pearson} and Appendix~\ref{Appendix1}), and the error bars are large (average of $\sim$50\,km\,s$^{-1}$). 
If we were only to account for the statistical errors from the continuum with the Monte-Carlo approach, we derive a tighter and statistically significant trend between these two parameters (slope 0.7$\pm$0.2; $\rho \sim 0.8$ and p-value\,$< 0.001$). 
There is a similar trend also in the selected SW integrated region (not for the NE, see Fig.~\ref{FigAp:v98W80}). A similar correlation was detected with MIRI data for the galaxy NGC\,7469, which was interpreted as a stratified outflow \citep{Armus2023}, and previously with Spitzer data in \cite{Dasyra2011}, which was interpreted as stratification of the gas in the AGN narrow line region (NLR), with the high excitation lines typically closer to the AGN \citep{Filippenko1984,Ho2009}. 

We do not detect any statistically significant trend for both W80 and v98 parameters with the critical density, n$_{crit}$, of the lines \citep[as in][]{Dasyra2011} except for W80 in the nuclear and the SW regions where there is a moderate correlation (see Appendix~\ref{Appendix1} and Table~\ref{Table:pearson}). This explains the differences in W80 measured between [Ne\,V] at 14.3 and 24.3\,$\mu$m in Fig.~\ref{Fig:W80V98} and Fig.~\ref{FigAp:v98W80} (see also Table~\ref{Table:W80}), although in the nuclear region they are still consistent within the errors. The critical density is larger for the [Ne\,V] at 14.3\,$\mu$m than at 24.3\,$\mu$m (log n$_{crit} =$\,4.51\,cm$^{-3}$ and 3.77\,cm$^{-3}$, respectively, see \citealt{Pereira-Santaella2022}), and the lines with high critical densities are produced more efficiently close to the AGN, that is also where the motions are faster and thus the kinematics could be more perturbed. These trends have been reported in other Seyferts using optical emission lines \citep[see e.g.][]{DeRobertis1986}.

We found that the largest values of the outflow velocity (v98\,$\sim$\,400-500\,km\,s$^{-1}$, maximum in absolute value between v02 and v98) are observed in the majority of the lines in the NE region (see Table~\ref{Table:W80}), thus indicating that this part is moving faster than the gas in the disc or the southern regions. This was expected from the profiles in the left panels of Fig.~\ref{Fig:IntProfiles}, and it is also seen in the velocity channel maps (see Figs.~\ref{FigAp:ChannelMaps},~\ref{FigAp:ChannelMaps_1}, and~\ref{FigAp:ChannelMaps_2}), and in the parametric modelling (see Sects.~\ref{SubSect3:Results_Disc} and~\ref{SubSect4:Discussion_Outflow}). We note that the average velocity of the lines to the NE (SW) are generally blueshifted (redshifted) with respect to the nuclear spectrum (see Table~\ref{Table:W80} and Fig.~\ref{Fig:IntProfiles}), although some nuclear lines have slightly blueshifted velocities as well. 

\begin{table}
\tiny
   \caption{Flux measurements of the main neon emission lines (R\,$=$\,0.5\arcsec).}
   \label{Table:fluxes}
   \centering          
   \begin{tabular}{lccccc}
		\hline\hline
		Line & IP & f$_{AGN}$ & f$_{NE}$ & f$_{SW}$ & f$_{C1}$\\ 
         & (eV) & \multicolumn{4}{c}{($\times 10^{-16}$ erg\,s$^{-1}$\,cm$^{-2}$)} \\
        \hline           
		$[$Ne\,II$]$  & 21.6  & 238.07$\pm$0.65 & 8.86$\pm$0.30 & 12.58$\pm$0.34 & 8.40$\pm$0.05 \\ 
        $[$Ne\,III$]$ & 41.0  & 377.35$\pm$0.64 & 7.07$\pm$0.34 & 12.82$\pm$0.14 & 5.68$\pm$0.09 \\ 
        $[$Ne\,V$]$   & 97.2  & 294.25$\pm$0.58 & 4.19$\pm$0.28 & 4.20$\pm$0.11  & 1.94$\pm$0.07 \\ 
        $[$Ne\,VI$]$  & 126.0 & 40.12$\pm$1.76  & 4.23$\pm$0.30 & 2.47$\pm$0.19  & --            \\
		\hline        
   \end{tabular}
   \tablefoot{The regions of interest are described in Sect.~\ref{Sect3:Results} and shown in Fig.~\ref{Fig:KinMaps_1comp}. All the fluxes have been estimated using a single Gaussian modelling. We have not applied an aperture correction.}
\end{table}

\subsection{Morphology of the ionised gas}
\label{SubSect3:Results_Morph}

The low-ionisation emission lines (IP\,$\lesssim$\,25\,eV) in this mid-infrared (mid-IR) range, such as [Ne\,II], [Ar\,II]\,6.98\,$\mu$m, [S\,III]\,18.7\,$\mu$m, and [Fe\,II]\,5.34\,$\mu$m (this line is discussed in Garc{\'i}a-Bernete et al. submitted), are typically produced by star formation (SF) processes \citep{PereiraSantaella2010b}, or shocks \citep{Allen2008}. The flux maps for these lines, in particular that of [Ne\,II] (see top left panel in Fig.~\ref{Fig:KinMaps_1comp}), show a disc-like emission oriented east-west, coincident with the direction of the disc in the galaxy (PA\,=\,88$^{\circ}$, AH23) and with the SF ring (1.1 to 1.4 kpc in diameter, i.e. $\sim$6-7\arcsec, AH23). The spatial resolution with ALMA in AH23 (beam $\sim 0.07\arcsec$) is better than that of MIRI/MRS for the [Ne\,II] ($\sim 0.25\arcsec$; \citealt{Argyriou2023}), thus the circumnuclear disc structure is not as well resolved as with CO(3-2).

Higher ionisation emission lines (IP\,$>$\,90\,eV) are most likely produced by AGN photoionisation \citep{Armus2007,PereiraSantaella2010b}. This is reflected in the flux maps of the [Ne\,V] and [Ne\,VI] lines, where no emission is detected in the disc. Rather, it is concentrated in the nucleus and in the north-south direction with an apparent two-sided X-shaped morphology, especially prominent in [Ne\,VI] (see two bottom rows in Fig.~\ref{Fig:KinMaps_1comp}). We note that particularly the [Ne\,V] intensity map could have some contamination from the PSF (see Sect.~\ref{SubSect3:Results_HighIonLines} and Fig.~\ref{FigAp:ContinuumMaps}). The bi-cone is extended to at least $\sim$2.5\arcsec\,N and $\sim$3.8\arcsec\,S (i.e. 450\,pc and 680\,pc, respectively, in projection) with an opening angle of $\sim$128$^{\circ}$, similar to that found through the optical [O\,III]$\lambda$5007\AA\,emission \citep{Thomas2017}. The northern part of the cone was not detected in the near-IR line [Si\,VI] (AH23) due likely to the dust lane crossing the galaxy (see Sect.~\ref{Sect1:Introduction}), and only detected at larger distances in optical wavelengths \citep{Thomas2017}. The emission lines with intermediate IP (25\,-\,90\,eV), such as [Ne\,III], [Ar\,III]\,8.99\,$\mu$m, and [S\,IV]\,10.51\,$\mu$m (see discussion on this line in Garc{\'i}a-Bernete et al. submitted), show intermediate properties and morphologies between the low and high excitation lines. This is expected as they are ionised by a mixture of SF processes and AGN photoionisation \citep{PereiraSantaella2010b,GarciaBernete2017}. As seen in the second panel from the top of Fig.~\ref{Fig:KinMaps_1comp}, the flux map of [Ne\,III] is more concentrated in the nuclear region, but still elongated towards both the disc and the bi-cone direction, with complex kinematic (see Sect.~\ref{SubSect3:Results_Disc} and Fig.~\ref{FigAp:ChannelMaps_1}).

\begin{figure*}
    \includegraphics[width=.7\columnwidth]{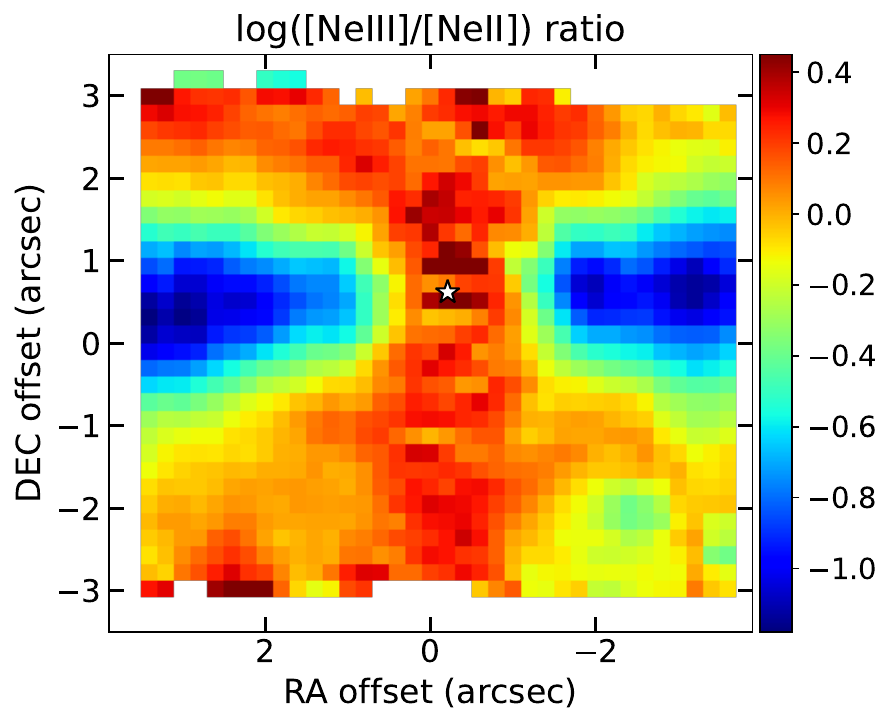}
    \includegraphics[width=.7\columnwidth]{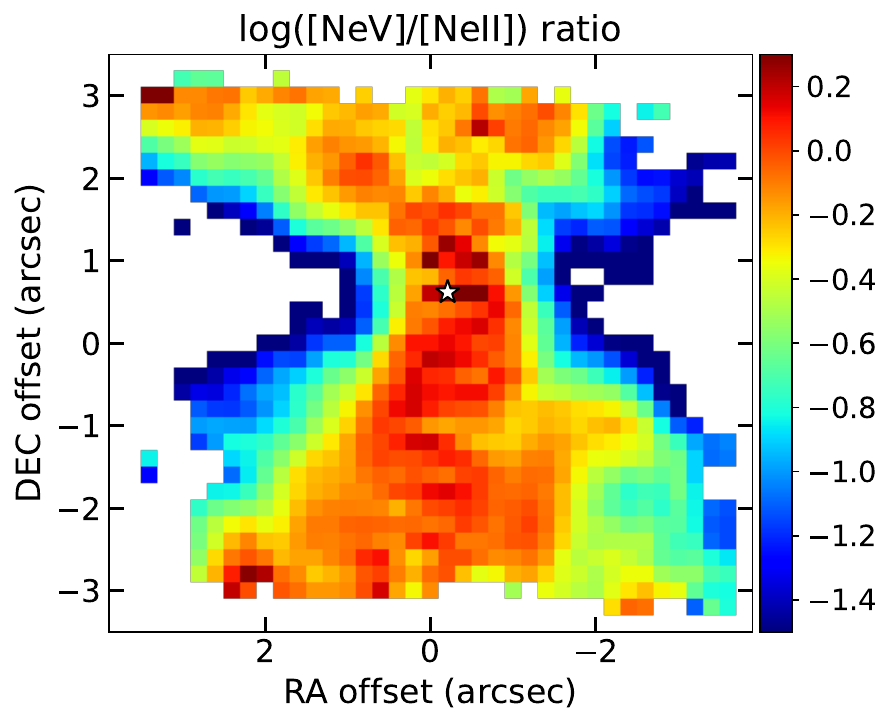}
    \includegraphics[width=.7\columnwidth]{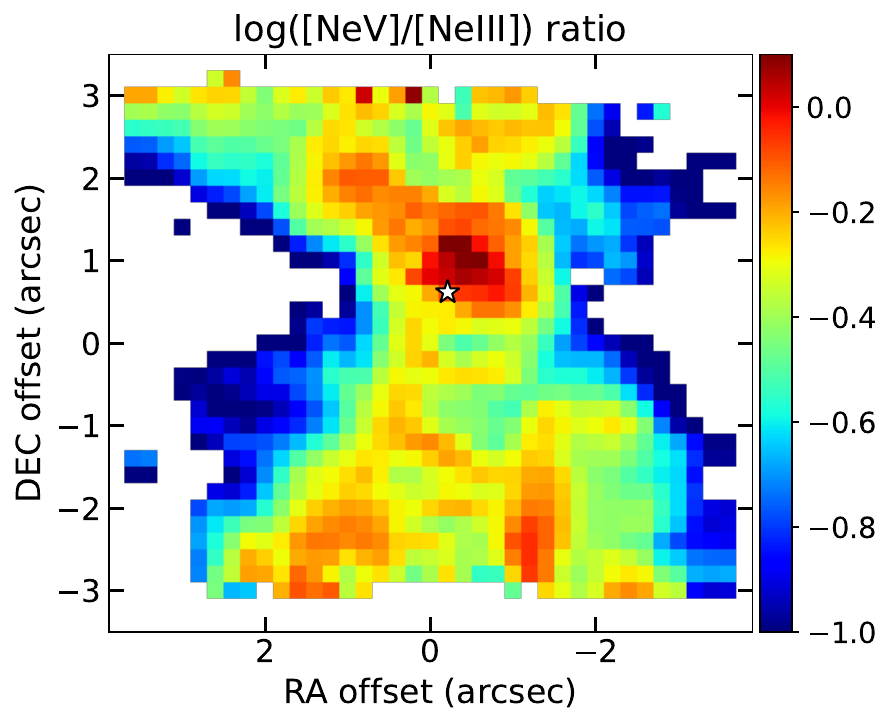}
     \caption{Maps of the [Ne\,III]/[Ne\,II], [Ne\,V]/[Ne\,II], and [Ne\,V]/[Ne\,III] line ratios (from left to right, respectively) represented in a logarithm scale. All the ratios are estimated after modelling the lines with a single Gaussian component (see Fig.~\ref{Fig:KinMaps_1comp}). The white star indicates the photometric centre of the flux maps.}
    \label{Fig:LineRatios}
 \end{figure*}

The [Ne\,III]/[Ne\,II] ratio map (Fig.~\ref{Fig:LineRatios}) shows a similar bi-conical structure extending in the north-south direction with increased line ratios with respect to the disc direction (average in log of $\sim$0.1 vs $\sim -$0.7, respectively; see left panel in Fig.~\ref{Fig:LineRatios}). These values are consistent with those of type-1 Seyferts and H\,II regions, respectively \citep[see Table~3 in][and Fig.~3 in Garc{\'i}a-Bernete et al. submitted]{PereiraSantaella2010b}. There is a distinct region located $\sim$3\arcsec\,SW from the nucleus with log([Ne\,III]/[Ne\,II])\,$\sim -$0.2 and log([Ne\,V]/[Ne\,II])\,$\sim -$0.7, that corresponds to one of the two kinematically-distinct clumps discussed in Sect.~\ref{SubSect3:Results_Disc} (see also Sect.~\ref{SubSect4:Discussion_Feedback}). The [Ne\,V]/[Ne\,III] ratio map (right panel in Fig.~\ref{Fig:LineRatios}) has several enhanced regions (log([Ne\,V]/[Ne\,III])\,$> -$0.2) mainly in the nuclear region and several knots in the ionisation cone. These values are consistent with the ratios derived for Seyfert galaxies using Spitzer/IRS data \citep{PereiraSantaella2010b}. 

\subsection{Spatially-resolved kinematics of the ionised gas}

\noindent In this section we present the results for the kinematics of the ionised gas emission lines according to their IPs (i.e. below and above 90\,eV). 

\subsubsection{Low excitation lines}
\label{SubSect3:Results_Disc}

The mean velocity field of [Ne\,II] (top-middle panel in Fig.~\ref{Fig:KinMaps_1comp}) shows a rotating pattern with maximum velocities $\sim$330\,km\,s$^{-1}$ in absolute value ($\Delta v$\,=\,$v_{\rm max} - v_{\rm min} \sim 530$\,km\,s$^{-1}$), oriented along the major axis of the disc (PA\,$\sim$\,88$^{\circ} \pm$\,1$^{\circ}$, measured north to east using the minimum and maximum velocities), with the receding side to the west. The velocity dispersion map (top-right panel in Fig.~\ref{Fig:KinMaps_1comp}) is not centrally peaked. Rather, the highest values are detected at $\sim$2\arcsec\,from the nucleus to both east and the west ($\sigma \sim$\,170\,km\,s$^{-1}$), as already seen in the cold molecular gas (AH23). When fitting the [Ne\,II] line with two Gaussians (see Fig.~\ref{FigAp:KinMaps_2comp}), the secondary component is mainly located along the disc direction, with a rotating pattern oriented as the primary component. This emission detected as a secondary component along the disc could be attributed to either the SF ring in the central parts, the presence of a nuclear bar, or both (see discussion in AH23). Additionally, the [Ne\,II] (top row in Fig.~\ref{Fig:KinMaps_1comp}) and also the [S\,III] intensity maps (see Fig.~\ref{FigAp:KinMaps_2comp}), show the presence of two distinct clumps, C1 and C2, at a projected distance of $\sim$3.1\arcsec\,and $\sim$4.3\arcsec\,(i.e. $\sim$\,560 and 780\,pc) with enhanced flux, low velocity dispersion ($\sigma \sim$\,50\,km\,s$^{-1}$) and blueshifted velocities. These are not detected in other low excitation lines because of their smaller FoVs (ch1 3.2\arcsec$\times$\,3.7\arcsec vs ch3 5.2\arcsec$\times$\,6.2\arcsec and ch4 6.6\arcsec$\times$\,7.7\arcsec). We estimated that the clumps have a size of R$\sim$0.5\arcsec\,(i.e. total size of $\sim$180\,pc), so they are resolved in the MRS data (PSF FWHM in ch3 varying from 0.56\arcsec\,to 0.65\arcsec, and 0.95\arcsec\,for ch4-short, see Fig.~\ref{Fig:KinMaps_1comp}). We focus on the results from the main clump at $\sim$\,540\,pc from the centre (C1) seen with [Ne\,II], as the C2 is not fully seen as it is located at the limits of the channel 3 FoV, and the [Ne\,II] line is detected with better spatial and spectral resolution than [S\,III]. We will further discuss these clumps in Sect.~\ref{SubSect4:Discussion_Feedback}. 

The ionised gas traced by intermediate IP lines, such as [Ne\,III] (see second row of Fig.~\ref{Fig:KinMaps_1comp}), is rotating along a PA ($\sim$80$^{\circ} \pm$\,2$^{\circ}$) and with a velocity amplitude similar to [Ne\,II] ($\Delta v \sim$\,490\,km\,s$^{-1}$). The velocity dispersion is irregularly distributed, non-centrally peaked with several regions with high $\sigma$ values (average $\sigma \sim$\,105\,km\,s$^{-1}$). The central, circular-like structure of the velocity dispersion indicates the presence of some contamination from the strong PSF emission (see ch3-long continuum map in Fig.~\ref{FigAp:ContinuumMaps} and \citealt{Argyriou2023}).
From the channel maps (see Fig.~\ref{FigAp:ChannelMaps_1}), it can be seen that the velocity is complex showing emission both in the disc and in the ionisation cone direction. The [Ne\,III] gas is ionised by a mixture of, on the one hand, SF processes which explains the similarities of the primary component to that of lines such as [Ne\,II] (see Fig.~\ref{Fig:KinMaps_1comp}), and on the other hand, AGN photoionisation that explains the detection of emission oriented south from the nucleus out of the disc, potentially related to the high-ionisation lines emission \citep[see e.g.][]{PereiraSantaella2010b}.

\subsubsection{High excitation lines}
\label{SubSect3:Results_HighIonLines}

The kinematic maps for the high ionisation emission lines are complex. The mean velocity field of the [Ne\,V] line (third row in Fig.~\ref{Fig:KinMaps_1comp}) fitted with one Gaussian shows mainly blueshifted velocities, whereas the velocity dispersion map is not homogeneous ($\sigma$ from 50 to 170\,km\,s$^{-1}$, with a standard deviation of 100\,km\,s$^{-1}$; see right panel, third row in Fig.~\ref{Fig:KinMaps_1comp}), with larger values north from the nucleus. The redshifted velocities ($<$\,50\,km\,s$^{-1}$) are mainly located in two regions NW and SW from the nucleus, whereas the most blueshifted velocities are found $\sim 3$\arcsec\,(i.e. $\sim$540\,pc) NE (see Fig.~\ref{Fig:KinMaps_1comp}). The intensity of the lines (see Fig.~\ref{Fig:KinMaps_1comp} and~\ref{FigAp:ChannelMaps_2}) seems to indicate that the east side of the cone (blueshifted) is more illuminated than the west side (redshifted). Moreover, the NE part of the cone is approaching at larger mean-velocities than the south-east part ($\sim -$350 vs $\sim -$250\,km\,s$^{-1}$), which suggest that the gas is expanding not completely parallel to the plane of the sky. We will further discuss the geometry of the system in Sect.~\ref{SubSect4:Discussion_Outflow}.

For [Ne\,VI] (bottom row in Fig.~\ref{Fig:KinMaps_1comp}), the emission is concentrated in an X-like shape oriented north-south from the centre (opening angle $\sim 90^{\circ}$), with the approaching (receding) side to the east (west) at a maximum mean-velocity of $\sim -200$\,km\,s$^{-1}$ ($\sim$80\,km\,s$^{-1}$). The velocity dispersion of the primary component is rather homogeneous (average $\sigma \sim 107$\,km\,s$^{-1}$), with higher values than the secondary component in the same region ($\sigma \sim$130\,km\,s$^{-1}$ vs $\sigma \sim$70\,km\,s$^{-1}$, respectively). 

The inner regions of the velocity and $\sigma$ maps for [Ne\,V] and [Ne\,VI] (see Fig.~\ref{Fig:KinMaps_1comp}) show the diffraction ring of the JWST PSF \citep[see][and continuum maps in Fig.~\ref{FigAp:ContinuumMaps}]{Argyriou2023}, which indicate that there is also an unresolved component in the line emission. Other galaxies in the GATOS sample, such as MCG-05-23-16 (see Esparza-Arredondo et al. in prep.), show unresolved intensity maps for [Ne\,V] and [Ne\,VI] lines, which are dominated by the PSF emission. For NGC\,7172, the petals from the PSF (see ch2-long and ch3-medium maps in Fig.~\ref{FigAp:ContinuumMaps}) are not clearly seen in the intensity maps (see Fig.~\ref{Fig:KinMaps_1comp}), thus the extended emission is likely not much affected by it.

We have also modelled these lines with multiple Gaussian components (see Sect.~\ref{Sect2:Data}), finding that the two-component modelling was the best preferred fit in several regions with similar properties for both lines. The second component concentrated mainly to the south and north-east of the AGN (see the modelling of [Ne\,V] in Fig.~\ref{FigAp:KinMaps_2comp_2}). There are blue and red wings everywhere along the bi-cone (see channel maps in Fig.~\ref{FigAp:ChannelMaps_2}), which would indicate that it is optically thin in these lines. Thus, the detection of various velocity components could be due to both the inclination of the cone with respect to the plane of the sky and its wide angle \citep[see also][]{Roy2021}. We describe the full properties of the two-component modelling in Appendix~\ref{Appendix2}.

\section{Discussion}
\label{Sect4:Discussion}

\subsection{The ionisation cone and the outflow}
\label{SubSect4:Discussion_Outflow}

The MIRI/MRS observations of NGC\,7172 have revealed for the first time a prominent biconical ionisation cone and associated outflow, which are particularly prominent for the [Ne\,V] and [Ne\,VI] lines (see Figs.~\ref{Fig:KinMaps_1comp} and~\ref{Fig:LineRatios}). The emission is observed in the north-south direction, reaching radial distances from the AGN $\sim$2.5\arcsec N and $\sim$3.8\arcsec S (i.e. projected distances of 450\,pc and 680\,pc, respectively). The ionisation cone extends further out than MIRI's FoV as seen from the [O\,III] emission in the southern part \citep{Thomas2017}. Despite the large extent of the outflow, NGC\,7172 has one of the lowest ionised gas mass outflow rates ($\dot{M}_{out} \sim$\,0.005\,M$_{\sun}$\,yr$^{-1}$) among the galaxies studied by \cite{Davies2020}. However, several effects might have affected the optical estimations of this parameter. 

The galaxy is crossed by a prominent dust lane (see Fig.~1 in AH23), which is obscuring the northern part of the ionisation cone as well as the nuclear emission even in the near-IR. Estimates of the optical extinction indicate A$_{V}> 3$\,mag \citep{Davies2020}. Particularly, based on the non-detection of the northern part in the [Si\,VI]1.96$\mu$m line in AH23, that region must be optically thick at that wavelength, which implies that the extinction at 2$\mu$m should be A$_{2\mu\,m} >$\,1\,mag \citep[or A$_{V} >$\,9\,mag][]{Wright2023}. This is corroborated by the extinction values of approximately A$_{V} =$\,15-20\,mag derived by \cite{Smajic2012} using near-IR hydrogen recombination lines. Moreover, both sides of the cone are similarly bright in [Ne\,V] and [Ne\,VI] (see Fig.~\ref{Fig:KinMaps_1comp}), which indicates that the extinction effects are not dominant for these lines in the mid-IR. This obscuration could have affected the measurements of the $\dot{M}_{out}$ for the ionised gas \citep[see Sect.~\ref{Sect1:Introduction} and][]{Davies2020} by a minimum factor of 5, plus at least a factor of two for the previous non-detection of the northern cone. 

The derived velocities for the outflow are modest (v98 up to $\sim$500\,km\,s$^{-1}$), but consistent with previous determinations in the optical \citep[see][and Fig.~\ref{Fig:W80V98}]{Davies2020}. These measurements however can be affected by the inclination of the cone with respect to our line of sight (see Sect.~\ref{SubSect3:Results_HighIonLines}), so to correct them a cone/outflow geometrical model would be needed. If the outflow axis was perfectly aligned with the plane of the sky, we would expect a symmetrical velocity structure. Instead, the outflow might be inclined such that the part tilted away from us is still within the line of sight, resulting in an overall blueshift in the average velocity maps. The wide aperture angle ($\sim 120^{\circ}$) could also contribute to the observed velocity structure. A significant portion of the outflow could have a velocity component towards us, even if the outflow is not significantly inclined. Additionally, the lack of redshifted emission could imply that the rear part of the bicone is obscured by the dust in the galaxy plane, and we are only seeing the front part due to its orientation. The ALMA detected torus of NGC\,7172 was proposed to be inclined at least $\sim$67$^{\circ}$ (AH23), so if the gas is launched from the inner regions in the perpendicular direction to the torus, it would explain the apparent asymmetry in the velocity maps. Assuming this geometry, the $\dot{M}_{out}$ would be affected by a factor of $\sim$2.4 due to the inclination of the outflow \citep[see Eq.~4 in][]{GarciaBurillo2014}.

The electron density, n$_{e}$, also affects the determination of $\dot{M}_{out}$. We estimated it from the ratios of the two [Ne\,V] lines using \textsc{pyneb} \citep{pyneb2015}. We measured the line fluxes in the nuclear integrated spectrum (R$\sim$1\arcsec), obtaining a ratio of f$_{\rm [Ne\,V]14}$/f$_{\rm [Ne\,V]24} = 1.52$, which is twice the value that was obtained by \cite{Spinoglio2015} with Spitzer data (integrating 4.7\arcsec\,for [Ne\,V]$_{14}$ and 11.1\arcsec\,for [Ne\,V]$_{24}$), but consistent with typical values for Seyfert galaxies \citep{PereiraSantaella2010b}. If we measure the fluxes at larger (smaller) radii, the ratio decreases (increases). Despite the fact that the mid-IR range is less affected by extinction, this galaxy has prominent silicate features in absorption that indicate that the emission could still be obscured \citep[see][and Garc{\'i}a-Bernete submitted]{GarciaBernete2024}. Following the extinction laws by \cite{Chiar2006} and \cite{Pei1992}, both [Ne\,V] lines fall in a minimum of the mid-IR extinction curve with similar extinction levels \citep[see also Fig.~4 in][]{HernanCaballero2020}. This means that the fluxes of these lines are equally, and not heavily, impacted by extinction. Therefore, the line ratio does not need to be corrected from extinction. 
We assumed a temperature of 10$^{4}$\,K for the gas, which gives n$_{e} \sim 3200$\,cm$^{-3}$, three times smaller than that estimated by \cite{Davies2020}. If we consider a R\,$=$\,2\arcsec\,instead, the ratio between the fluxes is 1.12 and the corresponding density would be $\sim$1150\,cm$^{-3}$, lower but in the same order of magnitude. We note that if we instead consider a range of temperatures from T$\sim$10$^{3}$ to 10$^{5}$\,K, the order of magnitude of n$_{e}$ remains between 10$^{3}$ to 10$^{4}$\,cm$^{-3}$ for the measured ratio, which is one order of magnitude larger than the estimates obtained using the [S\,II] emission lines, but consistent with the derivations using the ionisation parameter \citep[see][]{Davies2020}. Thus, considering a R of 1\arcsec\,to directly compare with the optical estimations, this would add a factor of 3 in the estimation of the ionised mass outflow rate. 

Taking into account all the different factors, the mass outflow rate for the ionised gas in NGC\,7172 might have been underestimated in \cite{Davies2020} using optical observations. Indeed, with the MIRI/MRS data Zhang et al. (submitted) followed \cite{Riffel2023} to estimate the mass outflow rate using the [Ne\,V]\,14$\mu$m, and derived a $\dot{M}_{out} \sim 0.03$\,M$_{\sun}$\,yr$^{-1}$ (not corrected by projection effects). This estimation assumed a spherical geometry for the outflow integrating with an aperture of 0.9\arcsec, to directly compare to that used to derive the optical value in \cite{Davies2020}. Thus the integrated value is six times larger than the previous measurement. Despite this, NGC\,7172 still remains as one of the galaxies from the GATOS sample with the lowest ionised mass outflow rate measurements (see detailed estimations of $\dot{\rm M}_{out}$ with the radius in Zhang et al. submitted.). We emphasise that this galaxy could be an extreme case among the GATOS sample because of the nearly edge-on orientation of the galaxy and the almost face-on orientation of the ionisation cone. It is likely that NGC\,7172 is a case of weak or moderate geometrical coupling \citep[see Fig.~4 in][]{RamosAlmeida2022}, where the expansion of the ionised gas outflow also triggers the expansion of the molecular gas producing another outflow in the direction of the disc, that was detected in CO with ALMA data in AH23. Thus in general, the ionised mass outflow rate of NGC\,7172 will be low compared to other galaxies with similar luminosities (see Zhang et al. submitted).

\subsection{Evidence for positive feedback}
\label{SubSect4:Discussion_Feedback}

We detected two kinematically distinct clumps exclusively in the [Ne\,II] (see C1 and C2 red circles in Figs.~\ref{Fig:KinMaps_1comp} and~\ref{FigAp:ChannelMaps}) and [S\,III] maps (see Fig.~\ref{FigAp:KinMaps_2comp}), located at $\sim$3.1\arcsec\,and $\sim$4.3\arcsec\,SW from the AGN, that are also detected in the MIRI narrow-band images (see Fig.~\ref{Fig:sketch}, Rosario et al. in prep). They were not seen in AH23 with ALMA data due to the smaller FoV in the north-south direction. The low velocity dispersion, distinct morphology in the [Ne\,II] flux map (see Sect.~\ref{SubSect3:Results_Morph}), and the observed line ratios (log([Ne\,III]/[Ne\,II])$\sim-0.3$; see Fig.~\ref{Fig:LineRatios}) suggest that they are SF regions. Moreover, the velocities are blueshifted with respect to the rotating primary component of the gas in the same spatial region (i.e. the primary component; see Sect.~\ref{SubSect3:Results_Disc} and Fig.~\ref{FigAp:ChannelMaps}). Given that the approaching side of the galaxy is the eastern part, these regions cannot be located in the plane of the galaxy but are likely close to the edge of the ionised gas cone/outflow. Emission from [Ne\,V] is also detected in the same region (see Fig.~\ref{FigAp:ChannelMaps_2}), but the flux map shows no distinct features at the clumps position as observed for [Ne\,II]. It is likely that the interaction between the expanding outflow and the interstellar medium (ISM) in the disc has entrained gas and triggered a SF episode. Using the [Ne\,II], [Ne\,III], and [Ne\,V] lines, we estimated a SF rate (SFR) for C1 of $\sim$0.08\,M$_{\sun}$\,yr$^{-1}$ using the relations by \cite{Zhuang2019}. As a comparison, we  estimated for an individual region of the circumnuclear ring (1.5\arcsec\,west from the nucleus, see Fig.~2 in AH23 and upper-left panel in Fig.~\ref{Fig:KinMaps_1comp}, integrating the fluxes with an aperture of R$\sim 0.5$\arcsec) a SFR of $\sim$1\,M$_{\sun}$\,yr$^{-1}$, which is 10 times larger than the SFR measured in the clumps. For comparison, in NGC\,92 the SFRs for those H\,II regions detected in the tidal tails range from 0.003 to 0.01\,M$_{\sun}$\,yr$^{-1}$ \citep{TorresFlores2009}. \cite{Cresci2015a} reported a case of positive feedback with the detection of two SF clumps, for which they derived a SFR $\sim$0.03\,M$_{\sun}$\,yr$^{-1}$, which is the same order of magnitude as the clumps in NGC\,7172.

\begin{figure}
    \includegraphics[width=\columnwidth]{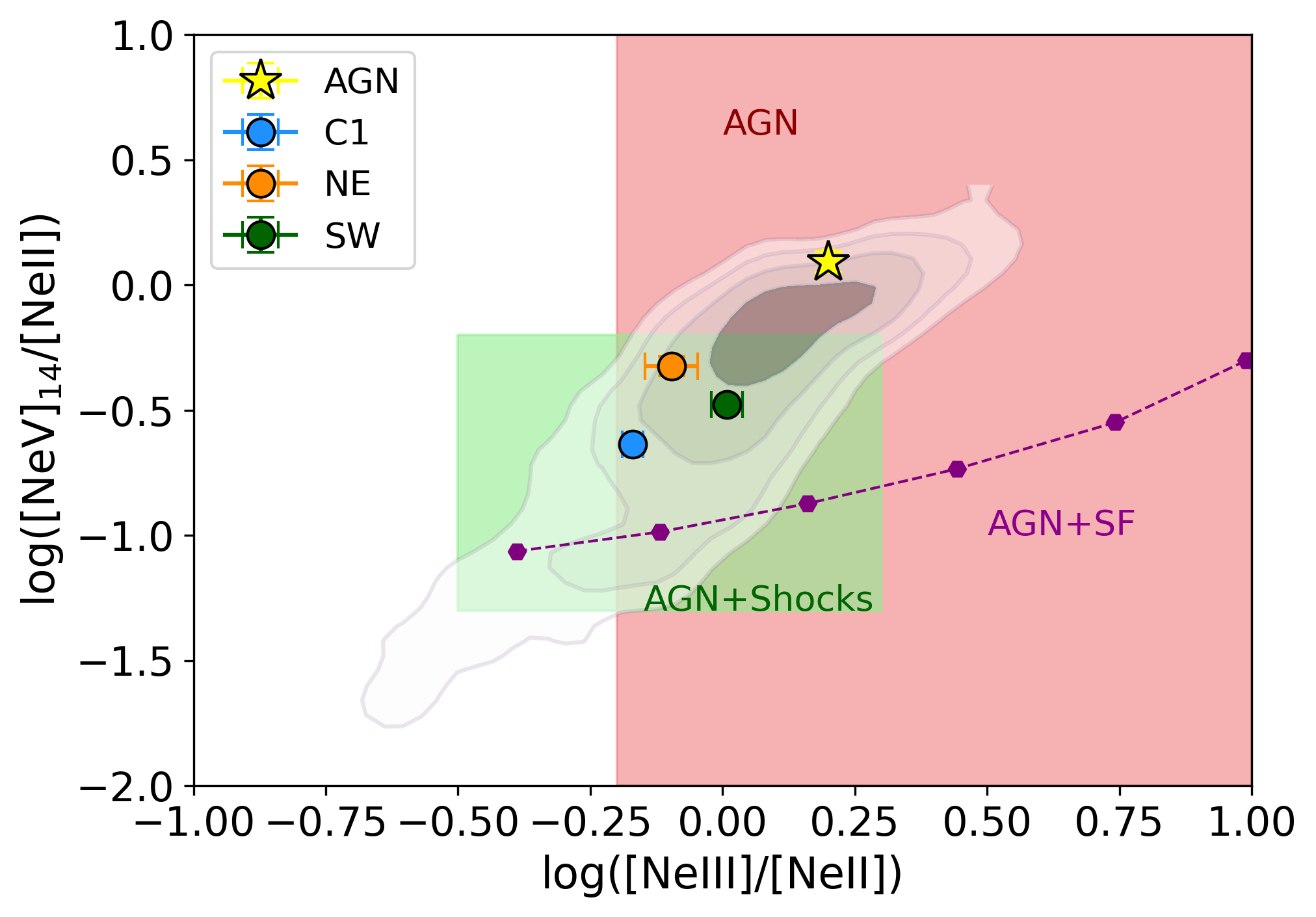}
     \caption{Mid-infrared line ratios following Fig.~5 in \cite{Feltre2023}. We represent the spatially resolved line ratios for NGC\,7172 as white and grey contours (levels from 1 to 10 spaxels). The yellow star represents the ratio estimated in the integrated spectrum for the nuclear region (AGN). The orange and green circles are the NE and SW regions in the cone; the blue circle represents the star-forming clump C1. The green region (i.e. square) represents AGN+shock models with velocities varying from 200 to 1000\,km\,s$^{-1}$; the red region represents AGN models at $\geq$\,1/3 solar metallicity; and the purple hexagons and line represents AGN+SF models with the ionisation parameter increasing from left to right (for more details, see Fig.~5 of \citealt{Feltre2023}).}
    \label{Fig:LineRatiosFeltre23}
 \end{figure}

\begin{figure*}
     \centering
     \includegraphics[width=\textwidth]{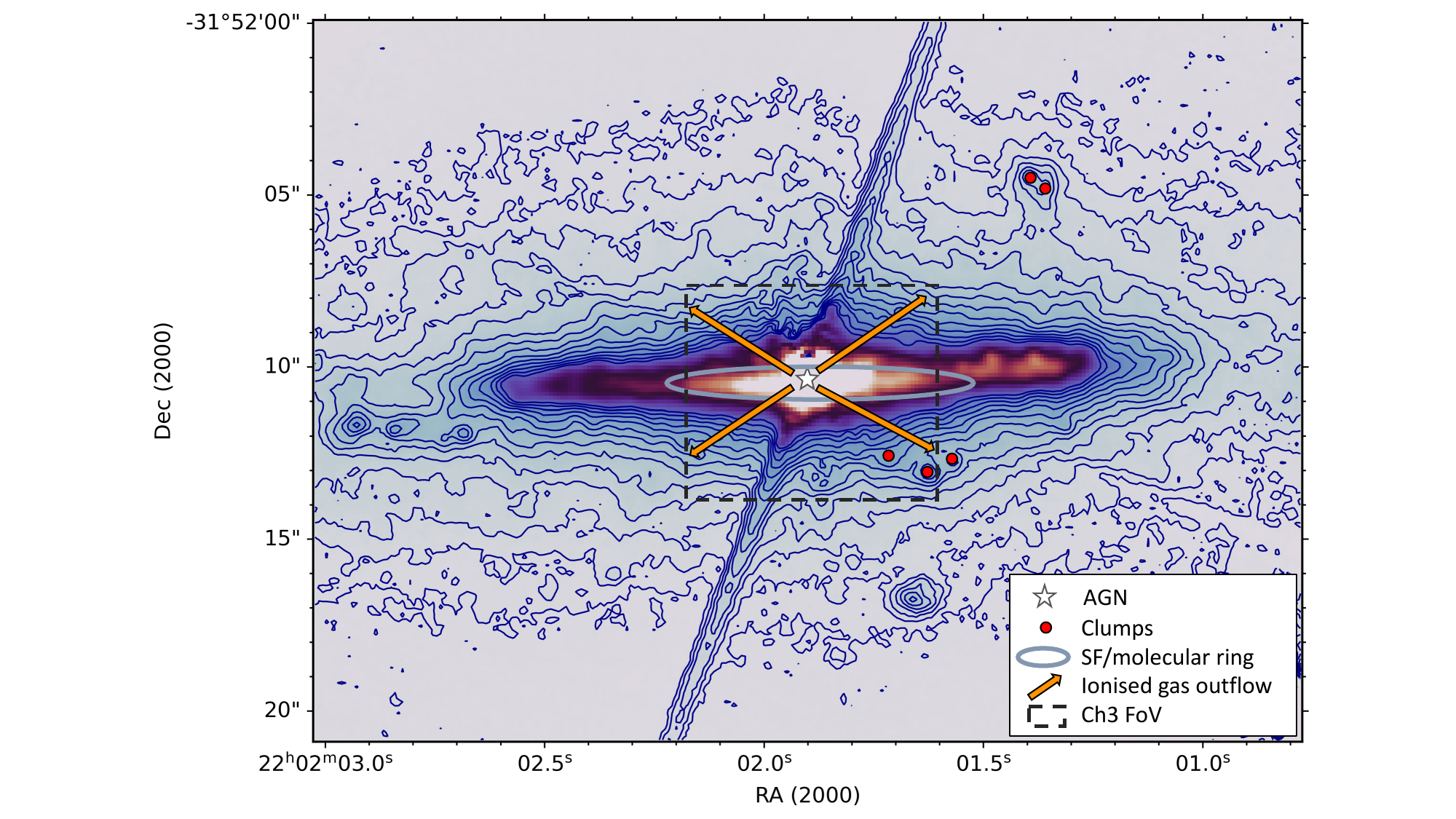}
     \caption{Schematic figure of the main structures detected in NGC\,7172 overplotted on the PSF-subtracted MIRI F1000W image from Rosario et al. in prep.  We indicate the aproximate size of the cold molecular gas ring detected with ALMA in red (see AH23), the AGN position with a white star, and the ionised gas outflow direction with yellow arrows (see Sect.~\ref{SubSect4:Discussion_Outflow}). Red circles indicate the position of several clumps detected in the MIRI image in the direction of the outflow, two of which coincide with the SF clumps seen with the MRS data (see top left panel of Fig.~\ref{Fig:KinMaps_1comp}, and Sects.~\ref{SubSect3:Results_Disc} and~\ref{SubSect4:Discussion_Feedback}). The bright emission crossing the central region from the SE to the NW across the image is a residual diffraction spike not fully eliminated when subtracting the bright PSF (see more details in Rosario et al. in prep.)} 
     \label{Fig:sketch}
  \end{figure*}

By qualitatively comparing the line ratios of these regions with the mid-IR diagnostic diagrams developed by \cite{Feltre2023}, the integrated value of the clumps (log([Ne\,III]/[Ne\,II])\,$\sim -0.11 \pm 0.02$ and log([Ne\,V]/[Ne\,II])\,$\sim -0.62 \pm 0.01$) is consistent with photoionisation by a mixture of AGN, SF processes and shocks\footnote{Note that these values are estimated using a single Gaussian modelling (see Table~\ref{Table:fluxes}), which means that both the disc and the clump components contribute to these line ratios.} (see Fig.~\ref{Fig:LineRatiosFeltre23}). As a comparison, the integrated value for the nuclear region lies in the region exclusively excited by the AGN emission, whereas the selected regions of the cone (see Sect.~\ref{Sect3:Results}) lie in the AGN and shock regions. Given that these regions are located at the edges of the outflow (see Figs.~\ref{Fig:KinMaps_1comp},~\ref{FigAp:ChannelMaps_2}, and~\ref{FigAp:ChannelMaps_3}), the ratios could be explained as gas photoionised by the AGN expanding from the centre and interacting with the ISM. Consistently with this scenario, the shocks in the clumps could have been produced due to the compression of the ISM gas by the outflow, eventually triggering the SF process.
This suggests that positive feedback is taking place in NGC\,7172, similarly to what was found by \cite{Cresci2015b} for the nearby Seyfert NGC\,5643. However, unlike this galaxy, the radio emission in NGC\,7172 is confined to the inner 2\arcsec\,\citep{Thean2000} and is not clearly connected with the SF clumps. In AH23 the power of the radio emission was estimated and proposed to be capable of producing the molecular gas outflow seen in the disc, so we cannot discard a contribution from that emission. \cite{Bessiere2022} reported a similar case of enhanced recent star formation at the edge of the approaching side of the co-planar ionised outflow of Mrk\,34.

Fig.~\ref{Fig:sketch} shows a schematic representation of the outflow expansion direction (see Sect.~\ref{SubSect4:Discussion_Outflow}) along with other relevant features in NGC\,7172, such as the SF circumnuclear ring (AH23) and the clumps superimposed to a MIRI image in the F1000W filter (after the nuclear PSF subtraction), which is part of another GATOS Cycle 1 program (PID: 2064, PI: D.~Rosario; Rosario et al. in prep.). We see in the same direction as clumps C1 and C2, there is another clump at $\sim 5$\arcsec\,from the AGN. This is located outside of the FoV of our MIRI/MRS observations. Other clumps are also visible to the NW in the MIRI image at larger projected distances ($\sim$7\arcsec) from the AGN. Other galaxies in the literature are known to have extraplanar H\,II regions \citep[see e.g.][]{DeRobertis1986,MendesOliveira2004,Stein2017} even at larger distances from the plane of the disc. Although there are various ways of producing these regions, such as ejections from the disc, the majority of them are typically explained through past or on-going interactions of the galaxies with other companions \citep[see e.g.][]{MendesOliveira2004,TorresFlores2009,MirallesCaballero2012}, so that the interaction triggers off-plane H\,II regions. NGC\,7172 is located in a group of galaxies \citep{Sharples1984} and the faint tidal tails seen in large-scale structure \citep[see Fig.~1 in AH23 and also][]{Sharples1984} are a clear signature of a past merger event. Thus we cannot rule out that some of the bright clumps detected in the MIRI image (see Fig.~\ref{Fig:sketch}) could have been formed during those processes. However, given that several clumps (i.e. C1, C2, and the third clump at 5\arcsec) are close (at least in projection) to the west edge of the outflow, they could be the product of a positive feedback episode.

\section{Summary and conclusions}
\label{Sect5:Conclusions}

We have presented the JWST MIRI/MRS observations of the GATOS galaxy NGC\,7172 to study the ionised gas kinematics with a particular focus on the neon emission lines (i.e. [Ne\,II], [Ne\,III], [Ne\,V], and [Ne\,VI]), that present a broad range of ionisation stages of the gas (i.e. IP from 20 to 130 eV). We have derived different kinematic properties of the lines through their modelling with parametric (i.e. one and two Gaussians per line) and non-parametric methods. Our results are as follows:

\begin{itemize}
    \item \textit{Low-intermediate ionisation emission lines:} The low-ionisation lines (e.g. [Ne\,II]) mainly trace the gas in the disc of the galaxy, with a secondary kinematic component located along the major axis probably related to the cold molecular gas ring detected with ALMA (AH23). Intermediate ionisation lines (e.g. [Ne\,III]) have a complex distribution both along the major and minor axis of the galaxy. They trace both gas emission and motions in the disc and part of the ionisation cone, as they are typically ionised by both the AGN emission and SF processes.
    \item \textit{High ionisation emission lines:} These lines (e.g. [Ne\,V] and [Ne\,VI]) are extended in the north-south direction from the centre out to the MIRI/MRS FoV edges (see Fig.~\ref{Fig:KinMaps_1comp} and~\ref{Fig:LineRatios}), and reveal a double-sided ionisation cone, which is oriented almost perpendicular to the galactic disc. Previous works had already detected an ionised gas outflow mainly with optical and near-IR data \citep{Thomas2017,Davies2020,AH2023}, but they were missing the northern cone probably due to the obscuration produced by the dust lane along the major axis of NGC\,7172. 
    \item \textit{Non-parametric kinematical modelling:} We found a mild correlation between the W80 parameter and the ionisation potential in the nuclear region ($\rho \sim 0.5$ and p-value\,$\sim 0.08$) and the SW region (projected distance of $\sim 360$\,pc) in the ionisation cone ($\rho \sim 0.65$ and p-value\,$\sim 0.03$). These correlations are typically interpreted as a stratification of the gas from the NLR, such that the high excitation lines are produced closer to the central parts. However in this case both the [Ne\,V] and [Ne\,VI] emission lines are extended up to the limits of MIRI/MRS's FoV ($\sim$\,3\arcsec, i.e. 540\,pc in projection).
    \item \textit{Properties of the outflow:} The bi-cone extends out to $\sim$2.5\arcsec\,N (i.e. 450\,pc) and $\sim$3.8\arcsec\,S (i.e. 680\,pc) with an opening angle $\sim$120$^{\circ}$, although potentially it could reach larger distances beyond MIRI/MRS's FoV based on its optical morphology \citep{Thomas2017}. The morphology, intensity, and modest outflow velocities (v98\,$\sim$400\,km\,s$^{-1}$) of the high ionisation lines suggest that the expansion of the cone is occurring not completely co-planar with the plane of the sky. With the MIRI data we are observing both the receding and the approaching sides of the cone. However, its orientation close to the plane of the sky means that the outflow velocities are only lower limits to the real velocities. We estimated that previous reported measurements of $\dot{M}_{out} \sim 0.005$\,M$_{\sun}$\,yr$^{-1}$ underestimated the ionised mass outflow rate mainly due to extinction and projection effects. Nevertheless, NGC\,7172 is likely a case of weak or moderate geometrical coupling, so it remains as one of the objects with lowest $\dot{M}_{out}$ in the GATOS sample (see also Zhang et al. submitted.).
    \item \textit{Feedback:} We detected two kinematically distinct clumps in the [Ne\,II] and [S\,III] maps at projected distances of $\sim 3.1\arcsec$ and $\sim 4.3\arcsec$ (i.e. 560 and 780\,pc) SW from the nucleus (see Figs.~\ref{Fig:KinMaps_1comp},~\ref{FigAp:KinMaps_2comp}, and~\ref{FigAp:KinMaps_2comp_2}), blueshifted with respect to the galaxy rotation, and thus likely taking part in the outflow. The observed [Ne\,III]/[Ne\,II] ratio however is typical of SF and shock regions (SFR\,$\sim 0.08$\,M$_{\sun}$\,yr$^{-1}$). Given their properties and location close to the SW edge of the cone (at least in projection), we propose that these regions may represent a case of positive feedback produced by the interaction of the expanding ionised gas outflow with the ISM of the galaxy. 
\end{itemize}
 
This work highlights the relevance of JWST data for studies on this kind, with the capability of detecting, characterising, and quantifying the properties of outflows in nearby AGN and study their effects on their host galaxies \citep[see e.g.][Zhang et al. submitted, Garc{\'i}a-Bernete et al. submitted]{Alvarez2023,Armus2023}. 

\begin{acknowledgements}
We thank the referee for his/her comments that have helped to improve the manuscript. LHM, AAH and MVM acknowledge financial support by the grant PID2021-124665NB-I00 funded by the Spanish Ministry of Science and Innovation and the State Agency of Research MCIN/AEI/10.13039/501100011033 PID2021-124665NB-I00 and by ERDF A way of making Europe. MPS acknowledges support from grants RYC2021-033094-I and CNS2023-145506 funded by MCIN/AEI/10.13039/501100011033 and the European Union NextGenerationEU/PRTR. IGB and DR acknowledge support from STFC through grants ST/S000488/1 and ST/W000903/1. SGB acknowledges support from the Spanish grant PID2022-1358560NB-I00, funded by MCIN/AEI/10.13039/501100011033/FEDER, EU. BGL and DEA acknowledge support from the Spanish Ministry of Science and Innovation through the Spanish State Research Agency (AEI-MCINN/10.13039/ 011000011033) through grants "Participation of the Instituto de Astrof{\'i}sica de Canarias in the development of HARMONI: D1 and Delta-D1 phases" with references PID2019-107010GB-100 and PID2022-140483NB-C21 and the Severo Ochoa Program 2020-2023 (CEX2019-000920-S). DEA also acknowledge financial support from MICINN through the programme Juan de la Cierva. DEA, AA, and CRA acknowledge funding from the State Research Agency (AEI-MCINN) of the Spanish Ministry of Science and Innovation under the grant "Tracking active galactic nuclei feedback from parsec to kiloparsec scales", with reference PID2022-141105NB-I00. MTL, CP, EKSH, and LZ acknowledge financial support from program JWST-GO-01670 provided by NASA through a grant from the Space Telescope Science Institute, which is operated by the Association of Unversities for Research in Astronomy, Inc., under NASA contract NAS 5-03127. EB acknowledges the Mar{\'i}a Zambrano program of the Spanish Ministerio de Universidades funded by the Next Generation European Union and is also partly supported by grant RTI2018-096188-B-I00 funded by the Spanish Ministry of Science and Innovation/State Agency of Research MCIN/AEI/10.13039/501100011033. AJB acknowledges funding from the "FirstGalaxies" Advanced Grant from the European Research Council (ERC) under the European Union’s Horizon 2020 research and innovation program (Grant agreement No. 789056). TDS acknowledges the research project was supported by the Hellenic Foundation for Research and Innovation (HFRI) under the "2nd Call for HFRI Research Projects to support Faculty Members \& Researchers" (Project Number: 3382). OGM acknowledges financial support from PAPIIT UNAM IN109123 and "Ciencia de Frontera" CONAHCyT CF-2023-G100. CR acknowledges support from Fondecyt Regular grant 1230345 and ANID BASAL project FB210003. MJW acknowledges support from a Leverhulme Emeritus Fellowship, EM-2021-064. MS acknowledges support by the Ministry of Science, Technological Development and Innovation of the Republic of Serbia (MSTDIRS) through contract no. 451-03-66/2024-03/200002 with the Astronomical Observatory (Belgrade).
This work is based on observations made with the NASA/ESA/CSA James Webb Space Telescope. The data were obtained from the Mikulski Archive for Space Telescopes at the Space Telescope Science Institute, which is operated by the Association of Universities for Research in Astronomy, Inc., under NASA contract NAS 5-03127 for JWST. These observations are associated with program 1670.
This research has made use of the NASA/IPAC Extragalactic Database (NED), which is operated by the Jet Propulsion Laboratory, California Institute of Technology, under contract with the National Aeronautics and Space Administration. 
This work has made extensive use of Python (v3.9.12), particularly with \textsc{astropy} \citep[v5.3.3, \nolinkurl{http://www.astropy.org};][]{astropy:2013, astropy:2018}, \textsc{lmfit} (v1.2.2), \textsc{pyneb} \citep[v1.1.18;][]{pyneb2015}, \textsc{matplotlib} \citep[v3.8.0;][]{Hunter:2007}, and \textsc{numpy} \citep[v1.26.0;][]{Harris2020}.
\end{acknowledgements}

%
   \bibliographystyle{aa} 
   \bibliography{bibliography.bib} 

%

\begin{appendix}

\section{Additional material of the non-parametric modelling}
\label{Appendix1}

In this appendix we show the properties of the emission lines from the integrated spectra in different regions estimated using a non-parametric methodology (see Sect.~\ref{Subsec3:Results_nonparam}). An example of the modelling of the integrated nuclear profile of [Ne\,V] line with both methodologies is shown in Fig.~\ref{FigAp:FitExamples}. On the one hand, with the non-parametric modelling we derive a W80\,$\sim$\,260\,km\,s$^{-1}$, v50\,$\sim -47$\,km\,s$^{-1}$, and a v98\,$\sim 306$\,km\,s$^{-1}$ (maximum in absolute value between v02 and v98). On the other hand, with the parametric modelling we used a single Gaussian with $\sigma \sim 114$\,km\,s$^{-1}$ (i.e. FWHM\,$\sim 268$\,km\,s$^{-1}$), and mean velocity $\sim -62$\,km\,s$^{-1}$. Thus for this integrated profile of [Ne\,V] the derived parameters are consistent with both methods. 
We note that we have made a cut of S/N\,$>3$\,at the peak of the line for measuring the properties of exclusively the ionised gas emission lines in Table~\ref{Table:W80}. Additional lines detected in the nuclear spectrum, such as molecular transitions, are reported in Zhang et al. (submitted) using a different methodology. 

As mentioned in Sect.~\ref{Sect3:Results}, for the nuclear region, we additionally applied a Monte-Carlo approach to estimate the uncertainty of the W80 measurement (i.e. measuring the v10 and v90 parameters) produced exclusively by the noise in the continuum. We varied the original continuum from the integrated spectra with random Gaussian noise, with $\sigma$ equal to the standard deviation of the original spectra. Then we changed the condition from defining the line as where the S/N was $>3$, to instead varying the line integration range from velocities in the range of $\pm$\,500 to $\pm$\,1000\,km\,s$^{-1}$ from the systemic velocity. We measured the v10 and v90 parameters for 30000 iterations for all the lines except for [Fe\,VIII] (only 5000 iterations) due to the strong features in the continuum. This approach allows to estimate the impact of the noise into the measurements, constraining the obtained W80 values. For the nuclear region the W80 values derived with both methods are consistent within the errors.

\begin{figure*}
    \includegraphics[width=0.67\columnwidth]{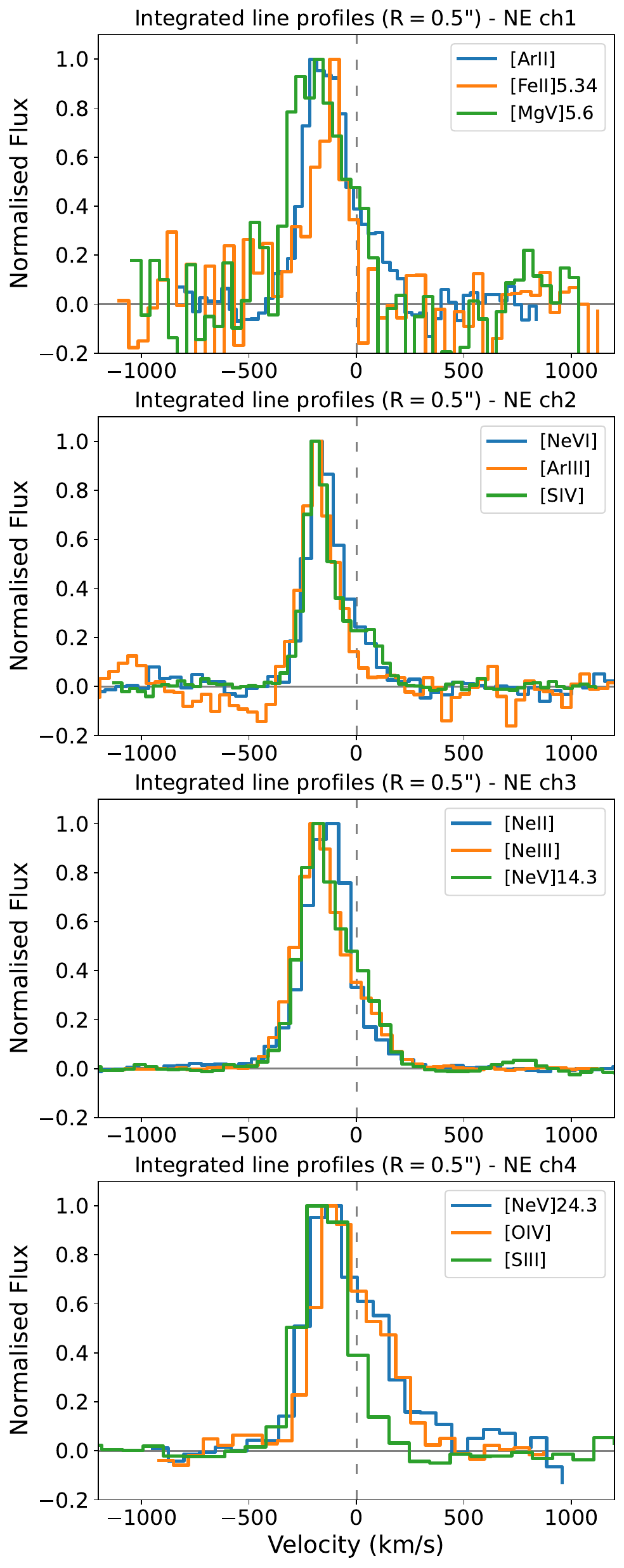}
    \includegraphics[width=0.67\columnwidth]{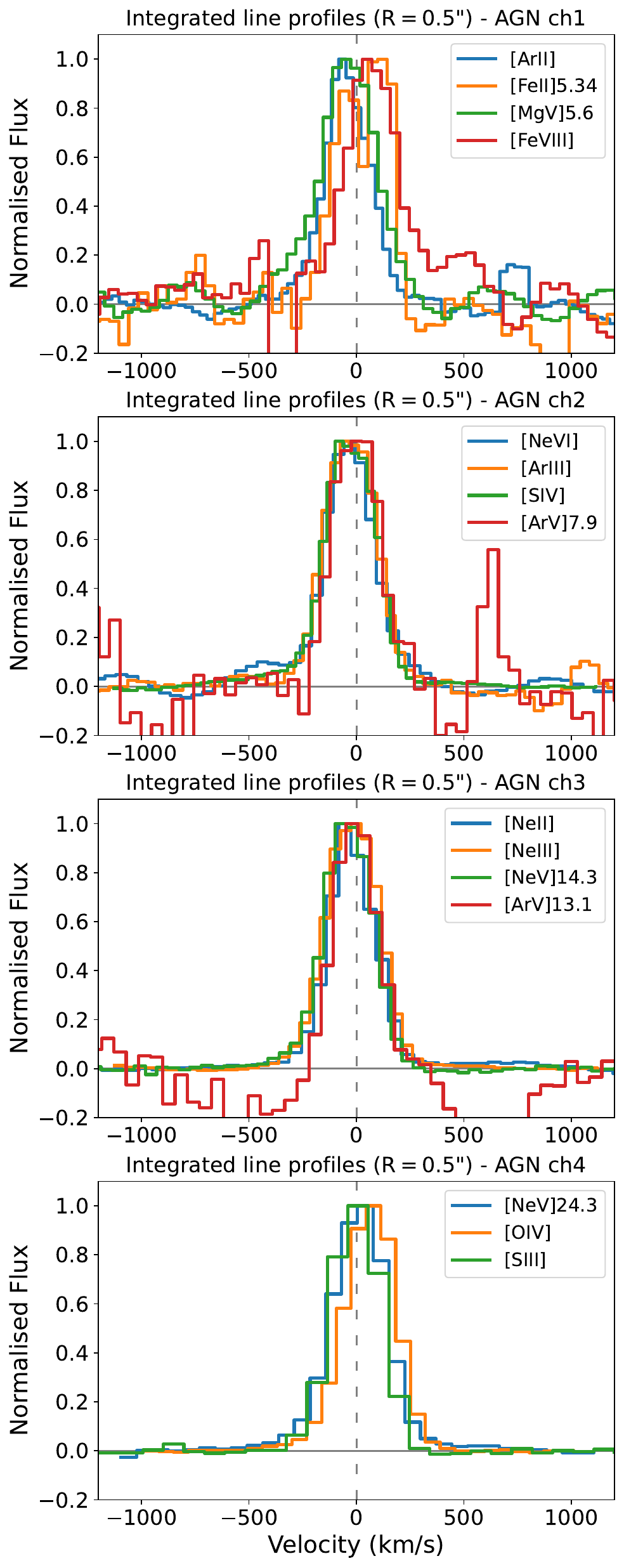}
    \includegraphics[width=0.67\columnwidth]{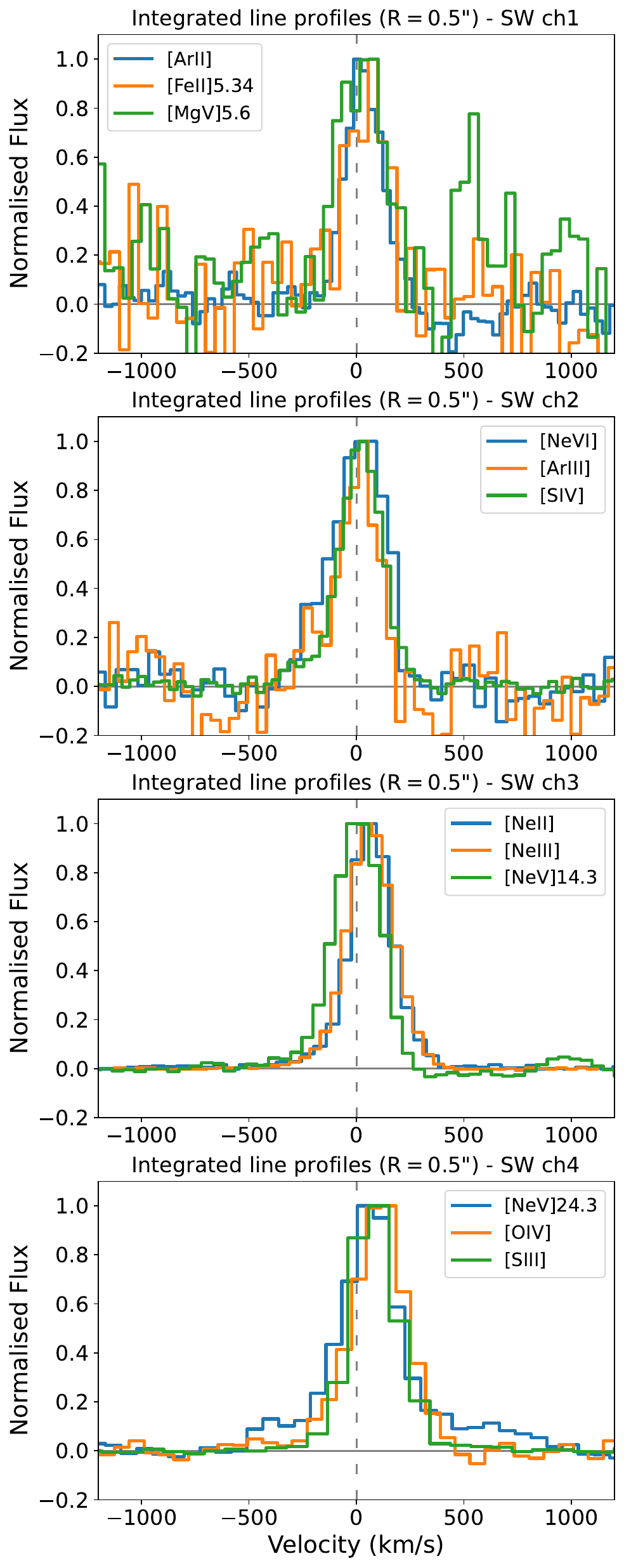}
     \caption{Same as Fig.~\ref{Fig:IntProfiles} but, from top to bottom, we show all the lines with S/N\,$>$\,3 distributed into the four individual channels of MIRI/MRS data (see Sect.~\ref{Sect2:Data}). All the velocities were corrected from the systemic value assuming a redshift of 0.00868 (see Sect.~\ref{Sect1:Introduction}).}
    \label{FigAp:IntProfiles_perchannel}
 \end{figure*}

\begin{figure*}
  \centering
    \includegraphics[width=\columnwidth]{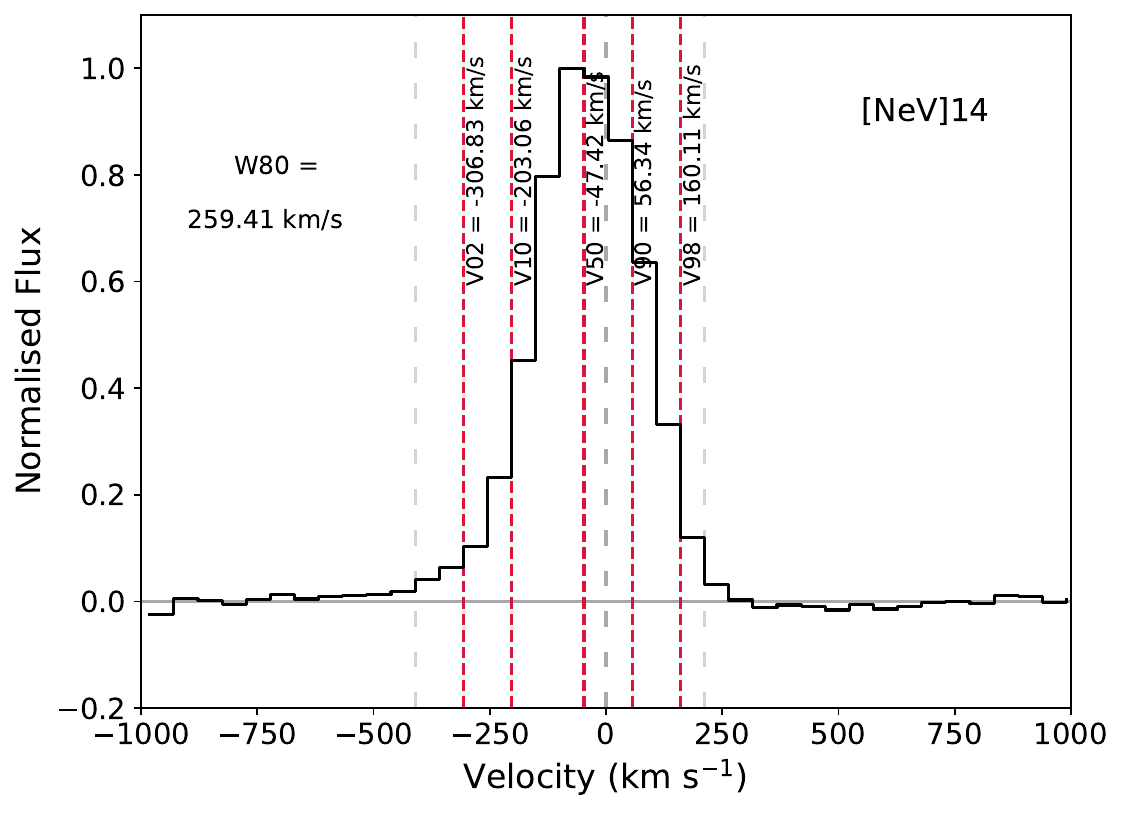}
    \includegraphics[width=.92\columnwidth]{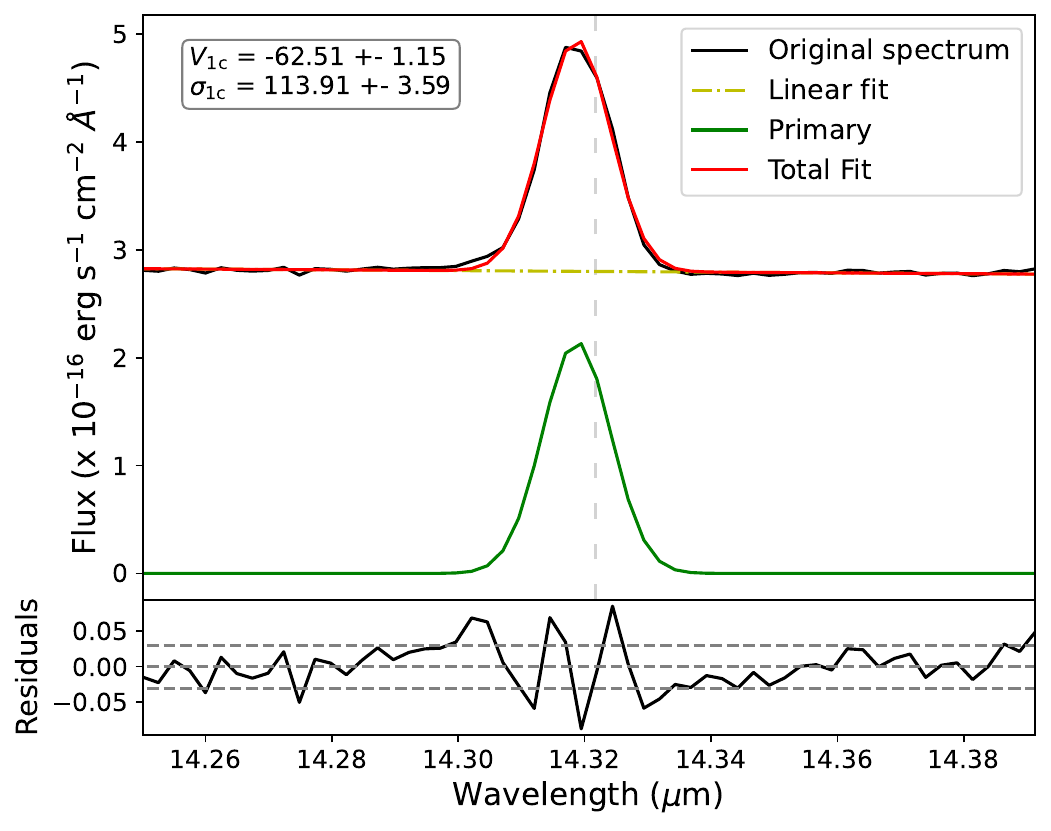}
     \caption{Examples on the modelling of the [Ne\,V]$_{14}$ line with the non-parametric (left) and the parametric techniques (right). \textit{Left panel:} Normalised flux of the line integrated in the nuclear region (R\,$= 0.5$\arcsec) with respect to the velocity. Grey, dashed, vertical lines indicate the limits of the line for doing the non-parametric modelling (see text in Appendix~\ref{Appendix1}). Red, dashed, vertical lines indicate the different parameters described in Sect.~\ref{Sect2:Data}. \textit{Right panel:} Gaussian modelling of the integrated nuclear region. The yellow line is the linear fit to the continuum, the green curve indicates the primary component, and the red curve indicates the global fit. The grey, dashed, horizontal lines in the residuals indicate the 3$\varepsilon$ limits (see Sect.~\ref{Sect2:Data}). The $\sigma$ is corrected from the instrumental value (see Sect.~\ref{Sect2:Data}). In both panels the central, grey, dashed, vertical line indicates the expected position of the line in rest frame, corrected from the systemic value with a redshift of 0.00868.}
    \label{FigAp:FitExamples}
 \end{figure*}

\begin{figure*}
  \centering
    \includegraphics[width=\columnwidth]{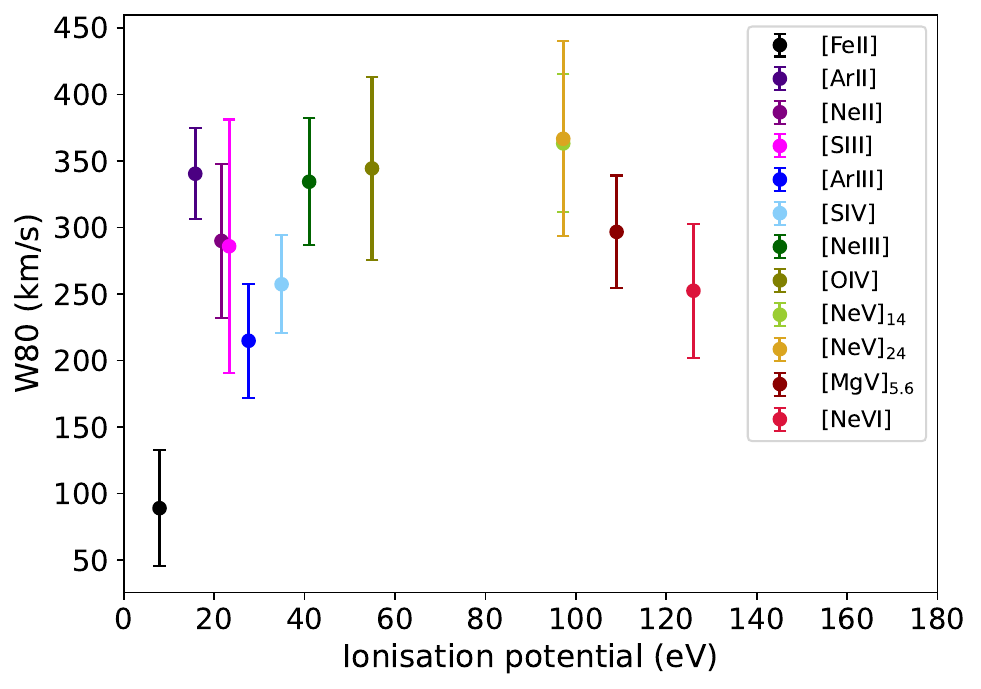}
    \includegraphics[width=\columnwidth]{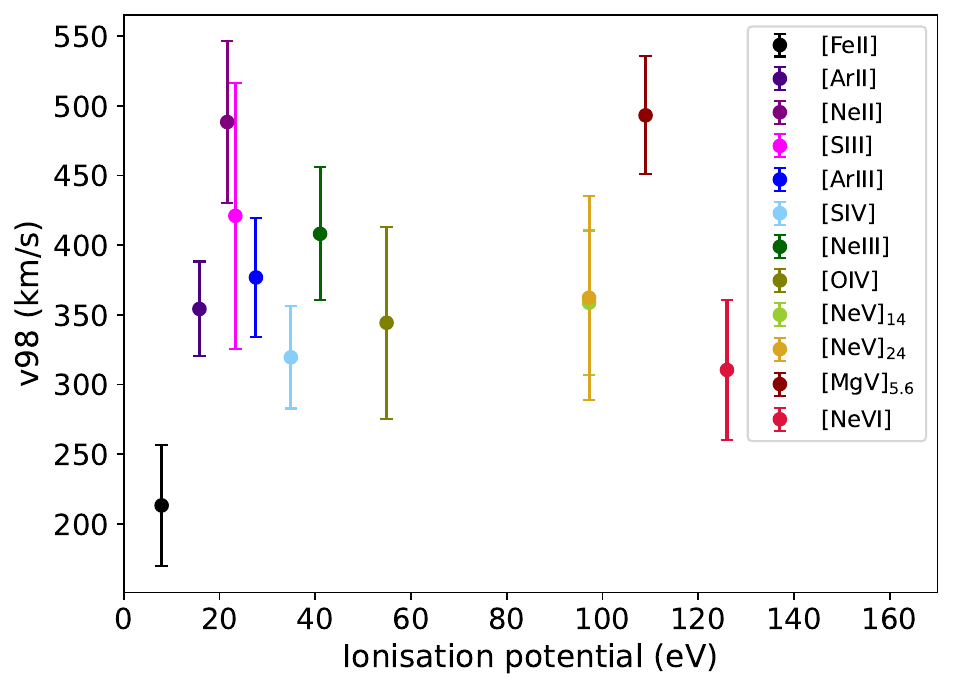}
    \includegraphics[width=\columnwidth]{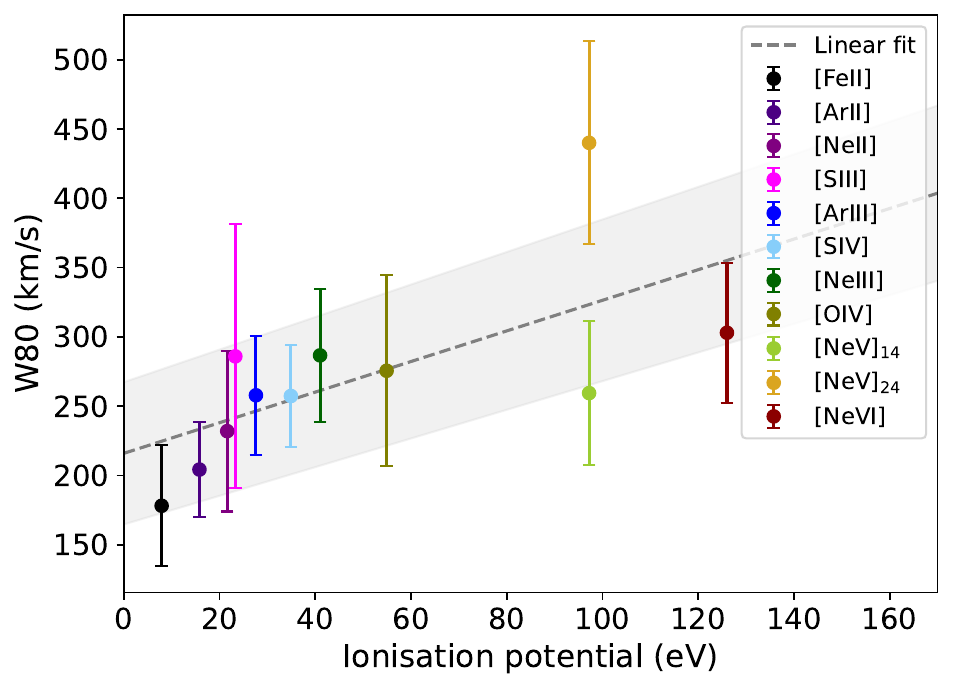}
    \includegraphics[width=\columnwidth]{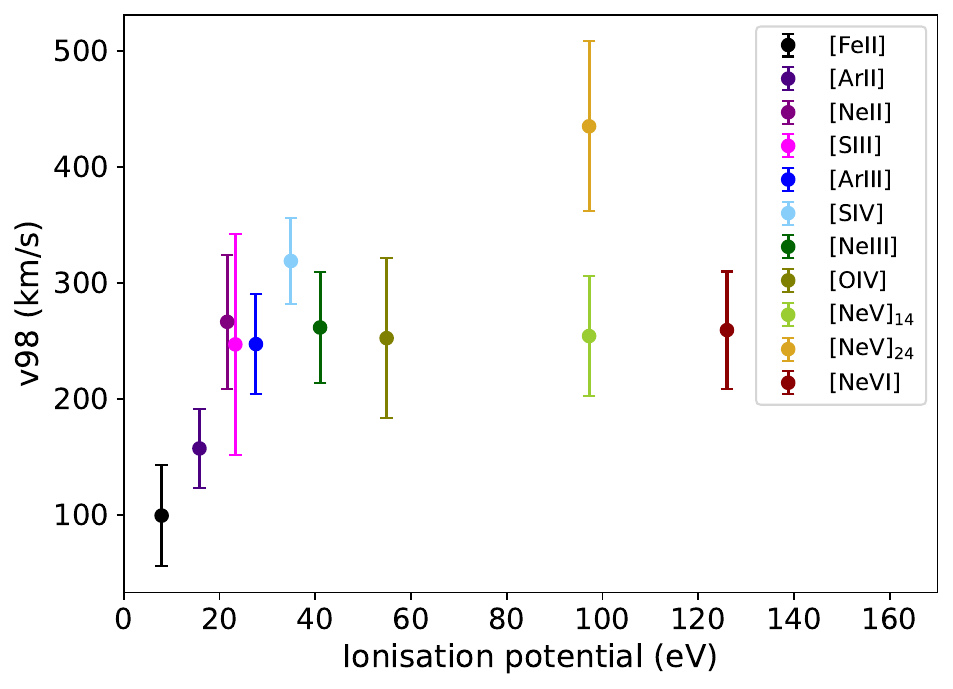}
     \caption{Same as Fig.~\ref{Fig:W80V98} but for the NE (top panels) and SW regions (bottom panels; see Sect.~\ref{Subsec3:Results_nonparam}).}
    \label{FigAp:v98W80}
 \end{figure*}

In general the integrated profiles in Fig.~\ref{FigAp:IntProfiles_perchannel} are symmetrical for the nuclear region (middle panels), and show redshifted and blueshifted wings for the two selected regions in the cone (NE and SW, respectively, see Sect.~\ref{Sect3:Results}). These wings are particularly prominent for the NE region, as can be seen in Table~\ref{Table:W80}, where the v98 parameter tends to be higher than for the nuclear region. 
For estimating the significance of the relation of these parameters with respect to the ionisation potential and the critical density, we computed the correlation coefficients $\rho$ and p-value (see Table~\ref{Table:pearson}). The Pearson coefficient, $\rho$, quantifies the strength of the relation between the variables, and the p-value indicates the probability of observing a correlation between the variables if there is no actual relation between them. None of the correlations are strong ($\rho > 0.7$), but rather moderate ($0.3 < \rho < 0.7$) for the nuclear aperture, and weak ($\rho < 0.3$) for the NE and SW apertures with some exceptions. We see a moderate correlation statistically significant (p-value\,$< 0.05$) for the W80 parameter with respect to the critical density in the nuclear region, and for the W80 parameter with respect to the IP in the SW. 
We did not detected any strong correlation of v98 with respect to the IP or the critical density of the lines in any of the regions (see Table~\ref{Table:pearson} and Figs.~\ref{FigAp:v98W80} and~\ref{Fig:W80V98}).  

\begin{table*}
   \caption{Measurements of the W80, v50, and v98 parameters of the ionised gas emission lines in the integrated nuclear spectra (R$\sim$0.5\arcsec).}
   \label{Table:W80}
   \centering          
   \begin{tabular}{lrrrrcccccc}
		\hline\hline
		Line & $\lambda$ & v50$_{\rm AGN}$ & v50$_{\rm NE}$ & v50$_{\rm SW}$ & W80$_{\rm AGN}$ & W80$_{\rm NE}$ & W80$_{\rm SW}$ & v98$_{\rm AGN}$ & v98$_{\rm NE}$ & v98$_{\rm SW}$ \\ 
		& ($\mu$m) & (km\,s$^{-1}$) & (km\,s$^{-1}$) & (km\,s$^{-1}$) & (km\,s$^{-1}$) & (km\,s$^{-1}$) & (km\,s$^{-1}$) & (km\,s$^{-1}$) & (km\,s$^{-1}$) & (km\,s$^{-1}$) \\ 
        \hline           
		$[\rm Fe\,II]$      & 5.34  &     9.3 & $-$168.8 &      -- & 267.1 & 89.1  & --    & 168.8 & 213.3 & --    \\ 
        $[\rm Fe\,VIII]$    & 5.45  & 27.5 &       -- &      -- & 305.6 & --    & --    & 289.4 & --    & --    \\ 
        $[\rm Mg\,V]$       & 5.61  & $-$69.4 & $-$196.6 &      -- & 339.1 & 296.7 & --    & 366.1 & 493.2 & --    \\ 
        $[\rm Ar\,II]$      & 6.98  & $-$47.9 & $-$150.1 &    21.0 & 272.3 & 340.4 & 204.2 & 252.2 & 354.3 & 217.3 \\
        $[\rm Ne\,VI]$      & 7.65  & $-$58.1 & $-$159.1 &  $-$6.7 & 302.9 & 252.5 & 252.5 & 411.5 & 310.5 & 259.2 \\ 
        $[\rm Ar\,V]$       & 7.90  & $-$24.0 &       -- &      -- & 342.3 & --    & --    & 219.6 & --   & --    \\ 
        $[\rm Ar\,III]$     & 8.99  & $-$33.1 & $-$250.0 & $-$32.2 & 300.8 & 214.9 & 257.8 & 248.0 & 376.9 & 248.0 \\
        $[\rm S\,IV]$       & 10.51 & $-$62.3 & $-$172.6 &    12.1 & 294.1 & 257.3 & 257.3 & 429.9 & 319.6 & 318.7 \\ 
        $[\rm Ne\,II]$      & 12.81 & $-$24.5 & $-$140.5 &    34.3 & 289.9 & 289.9 & 231.9 & 314.5 & 488.4 & 313.6 \\ 
        $[\rm Ar\,V]$       & 13.10 & $-$50.0 &       -- &      -- & 226.8 & --    & --    & 163.5 & --   & --    \\ 
        $[\rm Ne\,V]$       & 14.32 & $-$47.4 & $-$151.2 & $-$46.5 & 259.4 & 363.2 & 311.3 & 306.8 & 358.7 & 305.9 \\ 
        $[\rm Ne\,III]$     & 15.55 & $-$26.1 & $-$169.4 &    22.6 & 334.4 & 334.4 & 286.6 & 312.7 & 408.2 & 261.4 \\
        $[\rm S\,III]$      & 18.71 & $-$33.1 & $-$230.5 &    56.3 & 285.9 & 285.9 & 190.6 & 325.8 & 421.1 & 246.8 \\ 
        $[\rm Ne\,V]^{\dagger}$ & 24.32 & 4.3 & $-$142.4 &     5.2 & 366.7 & 366.7 & 440.0 & 362.4 & 362.4 & 434.8 \\ 
        $[\rm O\,IV]$       & 25.89 &    44.7 &  $-$93.1 &    45.6 & 344.4 & 344.4 & 344.4 & 299.7 & 344.4 & 321.1 \\
		\hline        
   \end{tabular}
   \tablefoot{$^{\dagger}$ indicates that the continuum near the line for the NE and SW regions contained strong wiggles and thus the measurements are not well constrained (see Fig.~\ref{FigAp:IntProfiles_perchannel}). The v98 parameter is the maximum in absolute value between the v02 and v98 measurements. The different integrated regions nuclear-AGN, NE, and SW, are defined in see Sect.~\ref{Sect3:Results}.}
\end{table*}

\begin{table}
\small
   \caption{Correlation coefficients between the W80 and v98 parameters with the lines ionisation potential and critical density at the three selected regions (see Sect.~\ref{Sect3:Results}). }
   \label{Table:pearson}
   \centering          
   \begin{tabular}{lcccccc}
		\hline\hline
		Relation & AGN &  & NE &  & SW &  \\ 
		& $\rho_{AGN}$ & p-value & $\rho_{NE}$ & p-value & $\rho_{SW}$ & p-value \\ 
        \hline           
		W80 vs IP             & 0.51 & 0.08 & 0.39 & 0.21 & 0.65 & 0.03  \\ 
        v98 vs IP             & 0.20 & 0.50 & 0.24 & 0.45 & 0.14 & 0.68  \\ 
        W80 vs log n$_{crit}$ & 0.69 & 0.01 & 0.13 & 0.70 & 0.54 & 0.09  \\ 
        v98 vs log n$_{crit}$ & 0.50 & 0.09 & 0.64 & 0.03 & 0.03 & 0.94  \\
		\hline        
   \end{tabular}
   \tablefoot{The Pearson coefficient, $\rho$, quantifies the strength of the relation between the variables, and the p-value indicates the probability of observing a correlation between the variables if there is no actual relation between them.}
\end{table}

\section{Additional spatially resolved maps}
\label{Appendix2}

\begin{figure*}
    \includegraphics[width=.69\columnwidth]{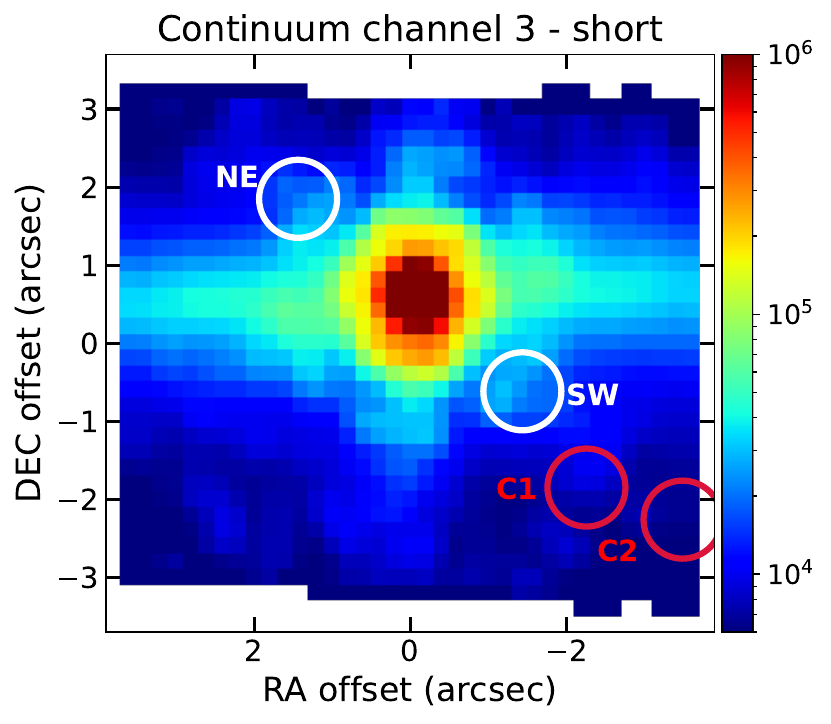}
    \includegraphics[width=.71\columnwidth]{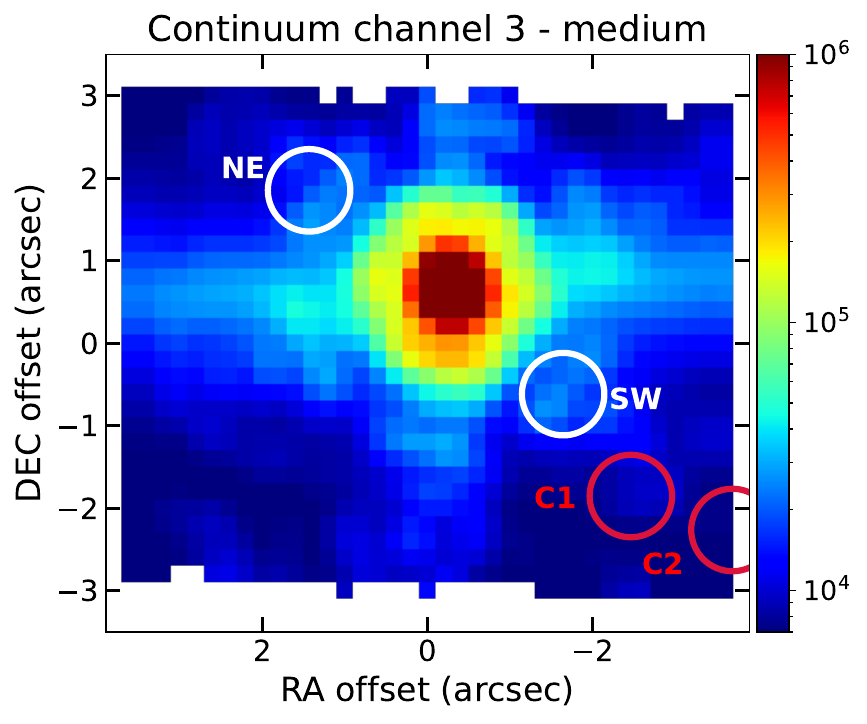}
    \includegraphics[width=.71\columnwidth]{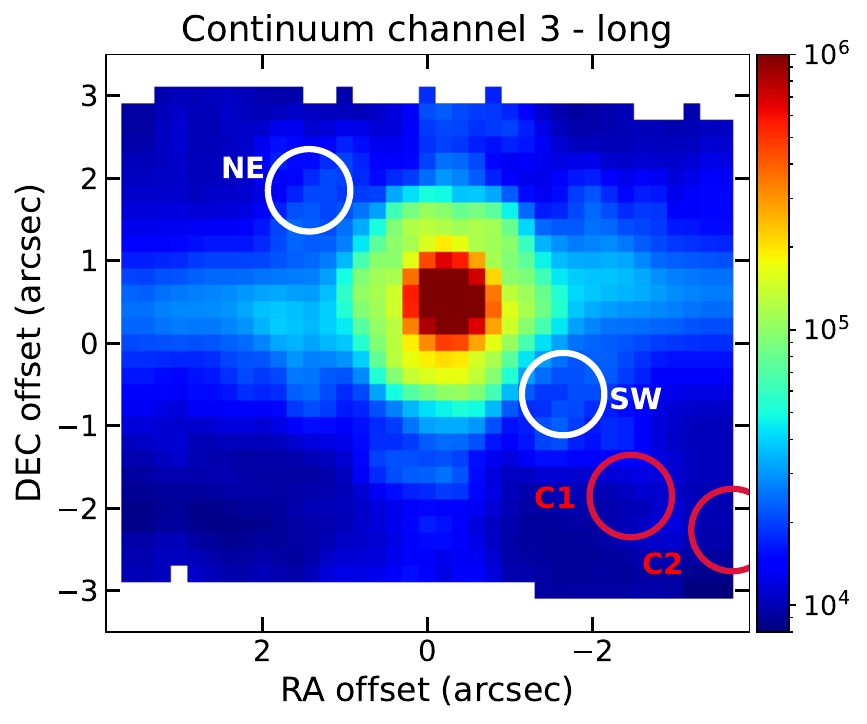}
    \includegraphics[width=.69\columnwidth]{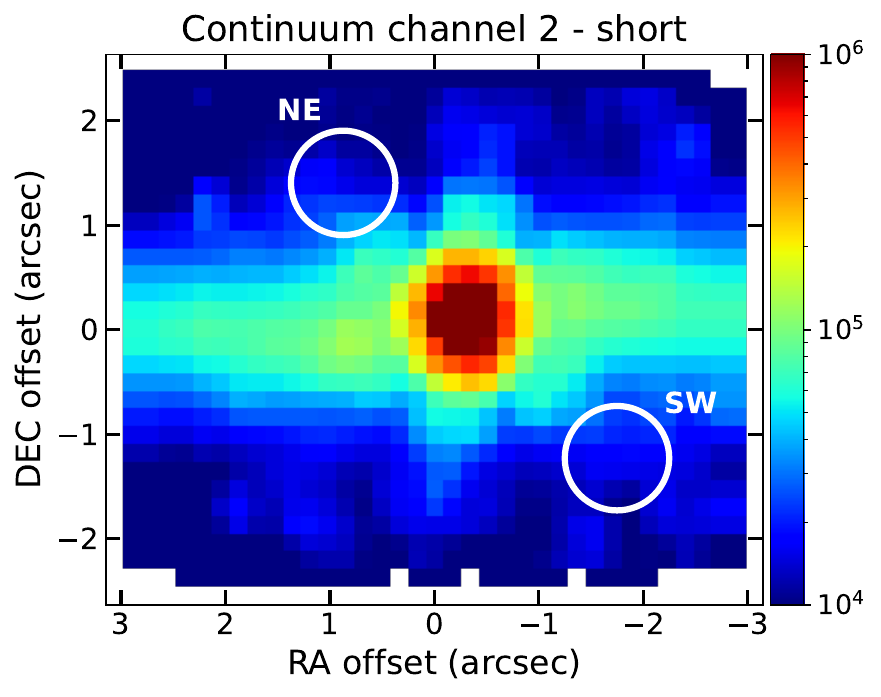}
     \caption{Continuum images for all the bands in channel 3 (upper panels from left to right, short, medium, and long, respectively) and the short band in channel 2 (bottom panel). The flux is in MJy\,sr$^{-1}$ and the regions correspond to those in Fig.~\ref{Fig:KinMaps_1comp}.}
    \label{FigAp:ContinuumMaps}
 \end{figure*}

\begin{figure*}
 \centering
    \includegraphics[width=0.95\textwidth]{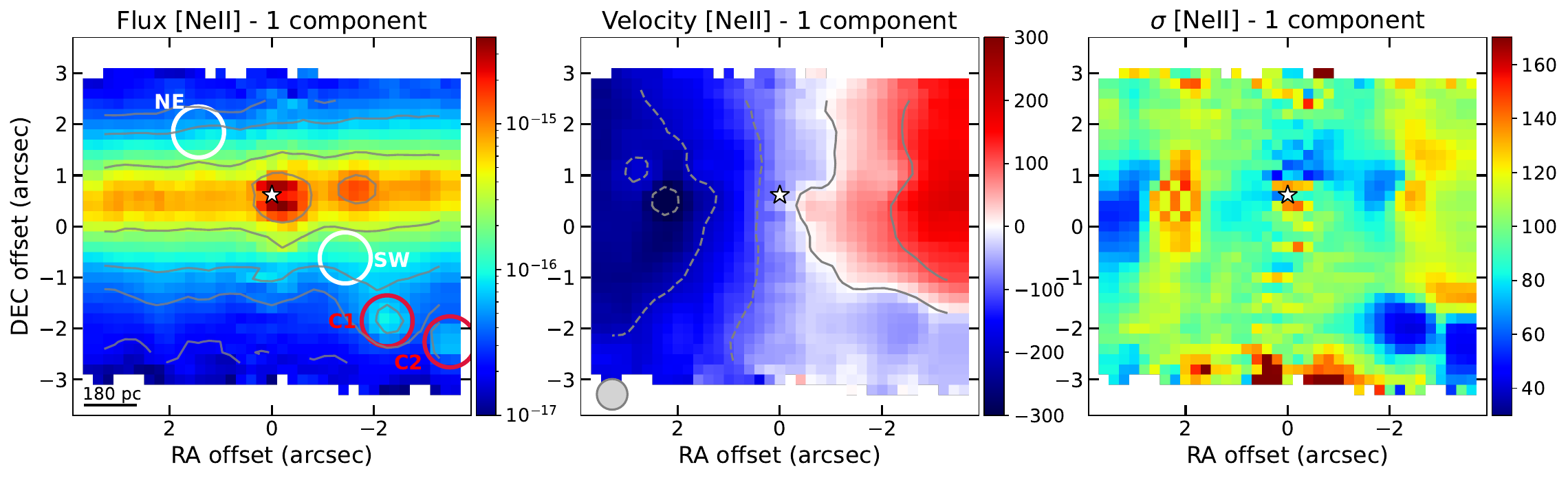}
    \includegraphics[width=0.95\textwidth]{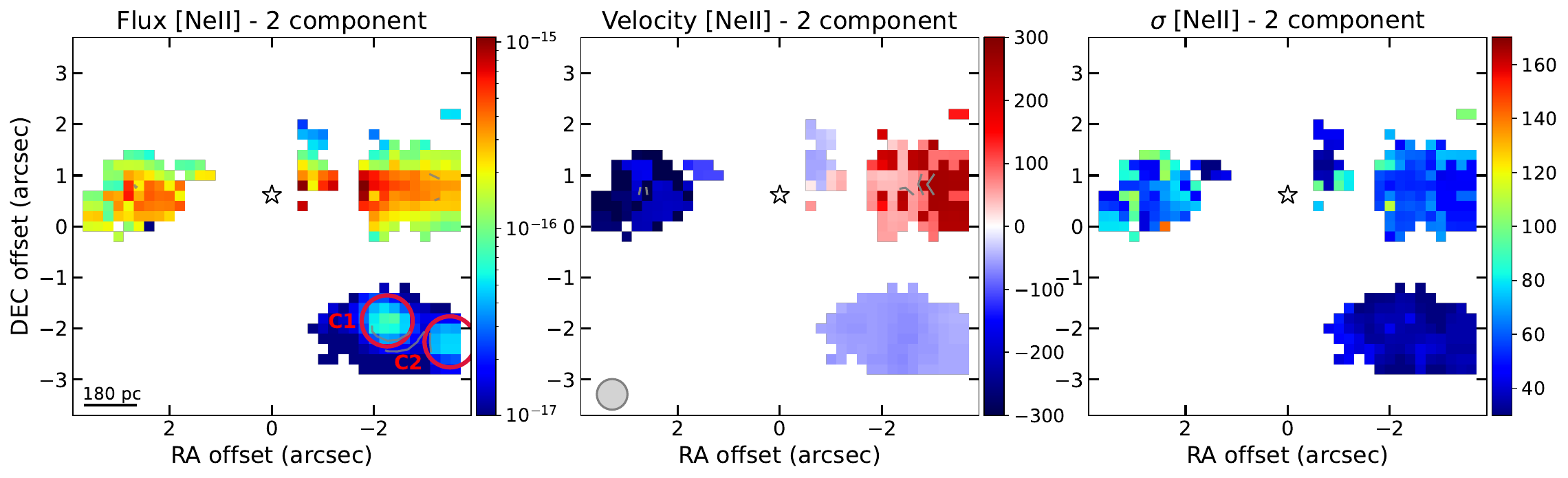}
    \includegraphics[width=0.95\textwidth]{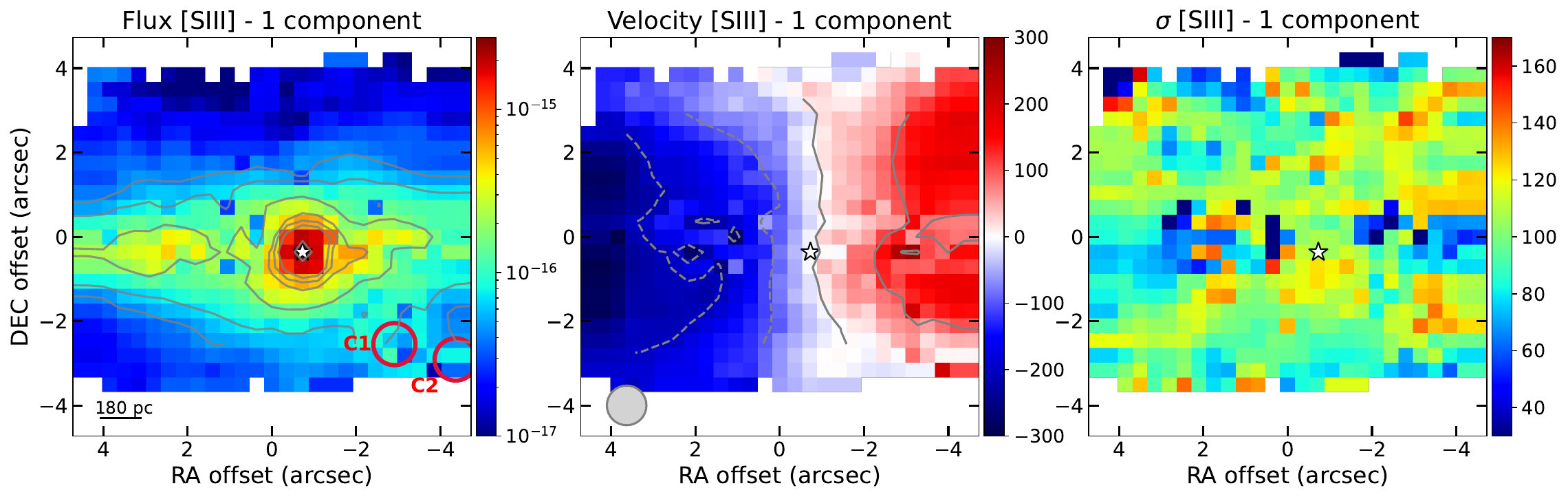}
    \includegraphics[width=0.95\textwidth]{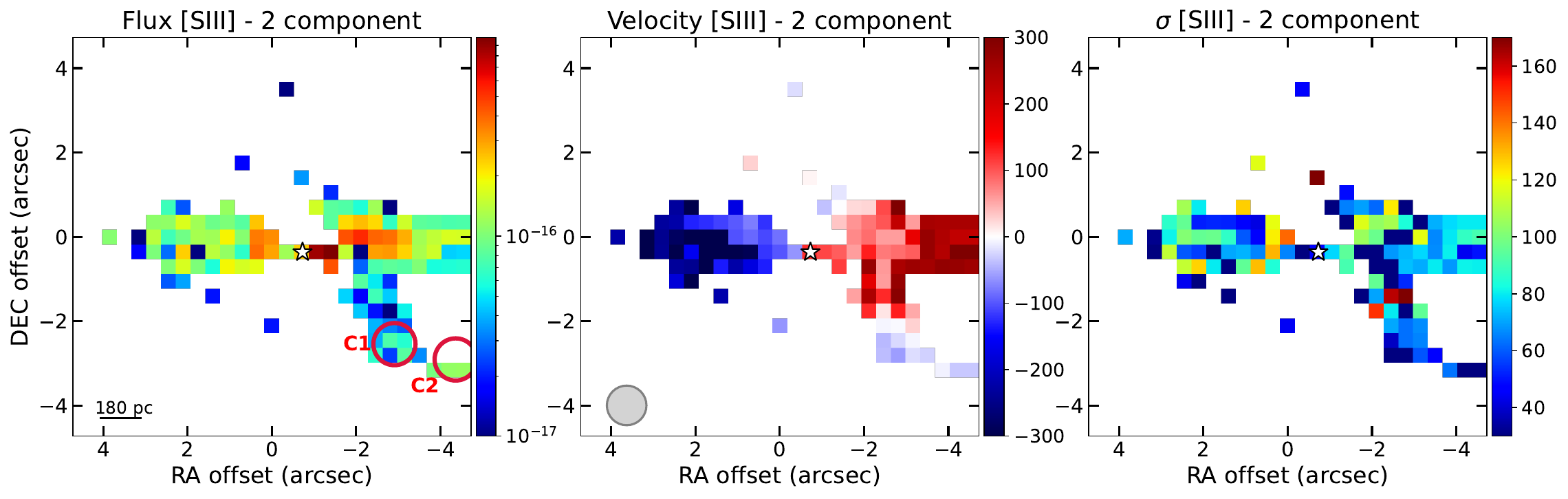}
    
     \caption{Maps of the two components model fit to the emission lines, following the description of Fig.~\ref{Fig:KinMaps_1comp}. From left to right we show the flux maps in erg\,s$^{-1}$\,cm$^{-2}$, the velocity, and the velocity dispersion in km\,s$^{-1}$. The photometric centre measured in the continuum of each channel is marked with a white star, the grey circle in the middle panels is the PSF, and the black line in the left panels is a 1\arcsec\,physical scale.} 
    \label{FigAp:KinMaps_2comp}
 \end{figure*}

\begin{figure*}
 \centering
    \includegraphics[width=0.95\textwidth]{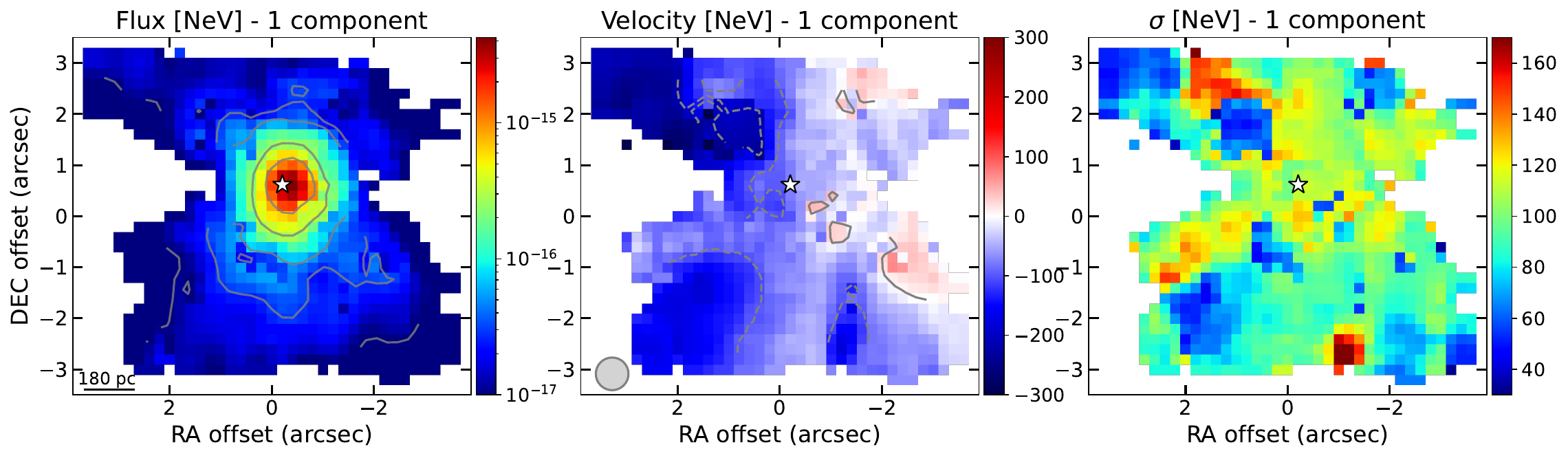}
    \includegraphics[width=0.95\textwidth]{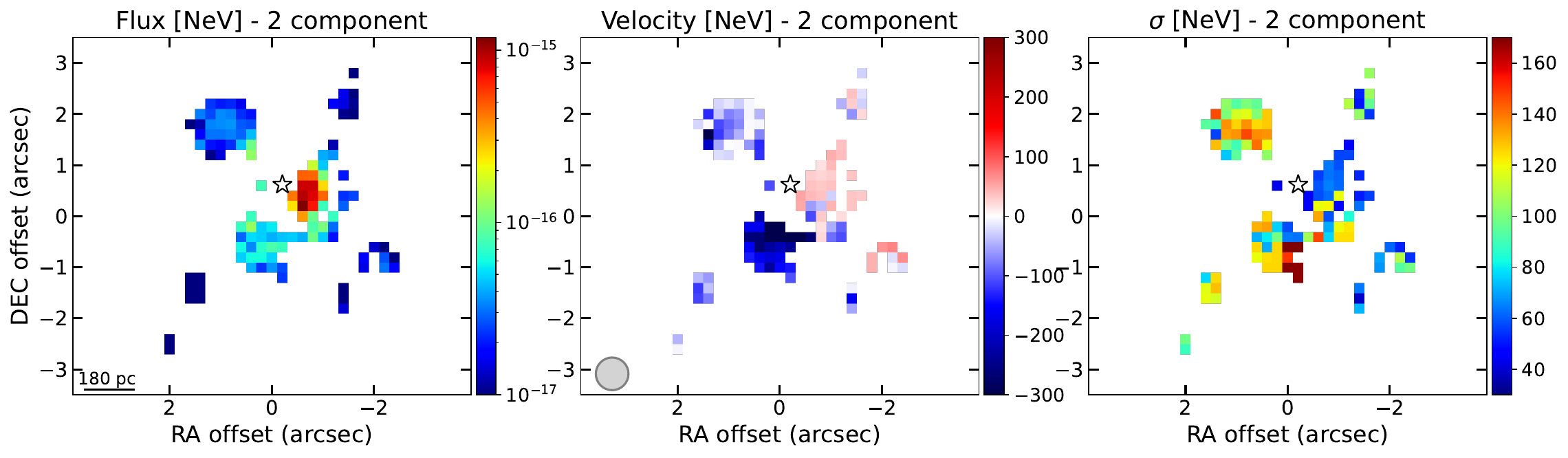}
     \caption{Same description as in Fig.~\ref{FigAp:KinMaps_2comp} but for [Ne\,V].}
    \label{FigAp:KinMaps_2comp_2}
 \end{figure*}

\begin{figure*}
 \centering
    \includegraphics[width=\textwidth]{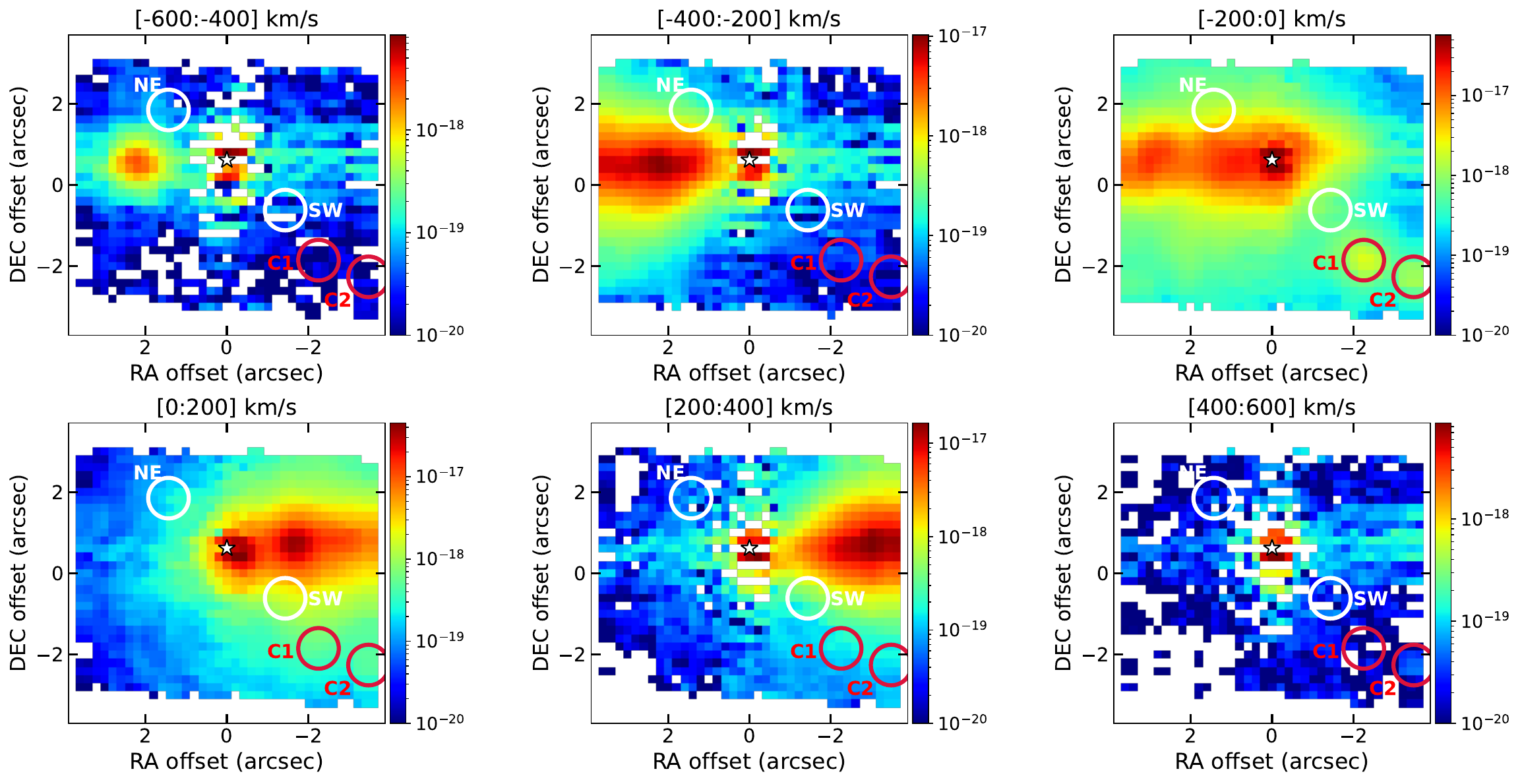}
     \caption{Channel maps of the [Ne\,II] line assuming steps of 200\,km\,s$^{-1}$. The white star is the same as Fig.~\ref{Fig:KinMaps_1comp}. White circles are the regions in which we obtained the spectra for the NE and SW parts of the ionisation cone, and the red circle is the main clump spectra (see Sect.~\ref{Sect3:Results}).}
    \label{FigAp:ChannelMaps}
 \end{figure*}

\begin{figure*}
 \centering
    \includegraphics[width=\textwidth]{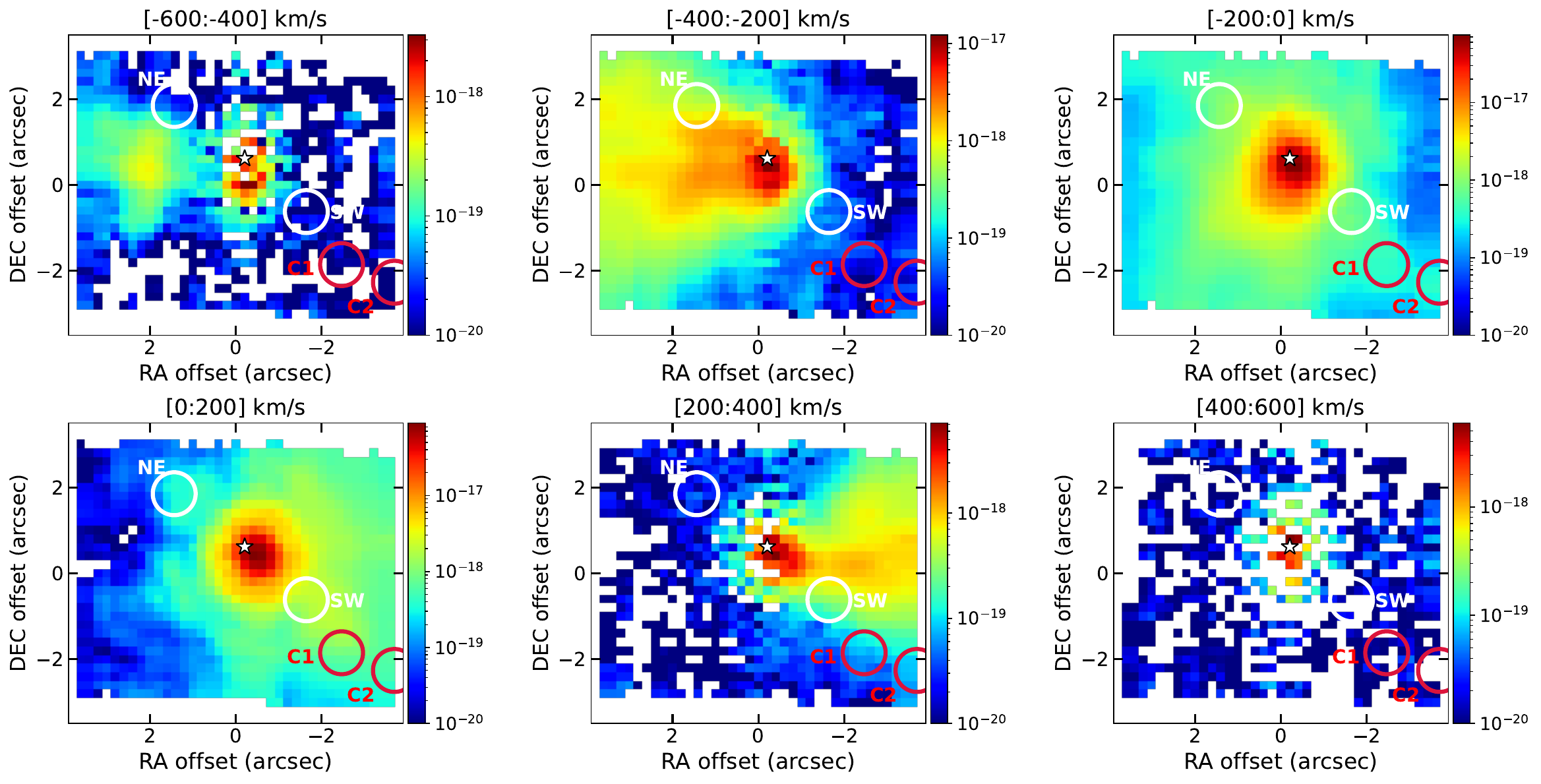}
     \caption{Same as Fig.~\ref{FigAp:ChannelMaps} but for [Ne\,III] line.}
    \label{FigAp:ChannelMaps_1}
 \end{figure*}

\begin{figure*}
 \centering
    \includegraphics[width=\textwidth]{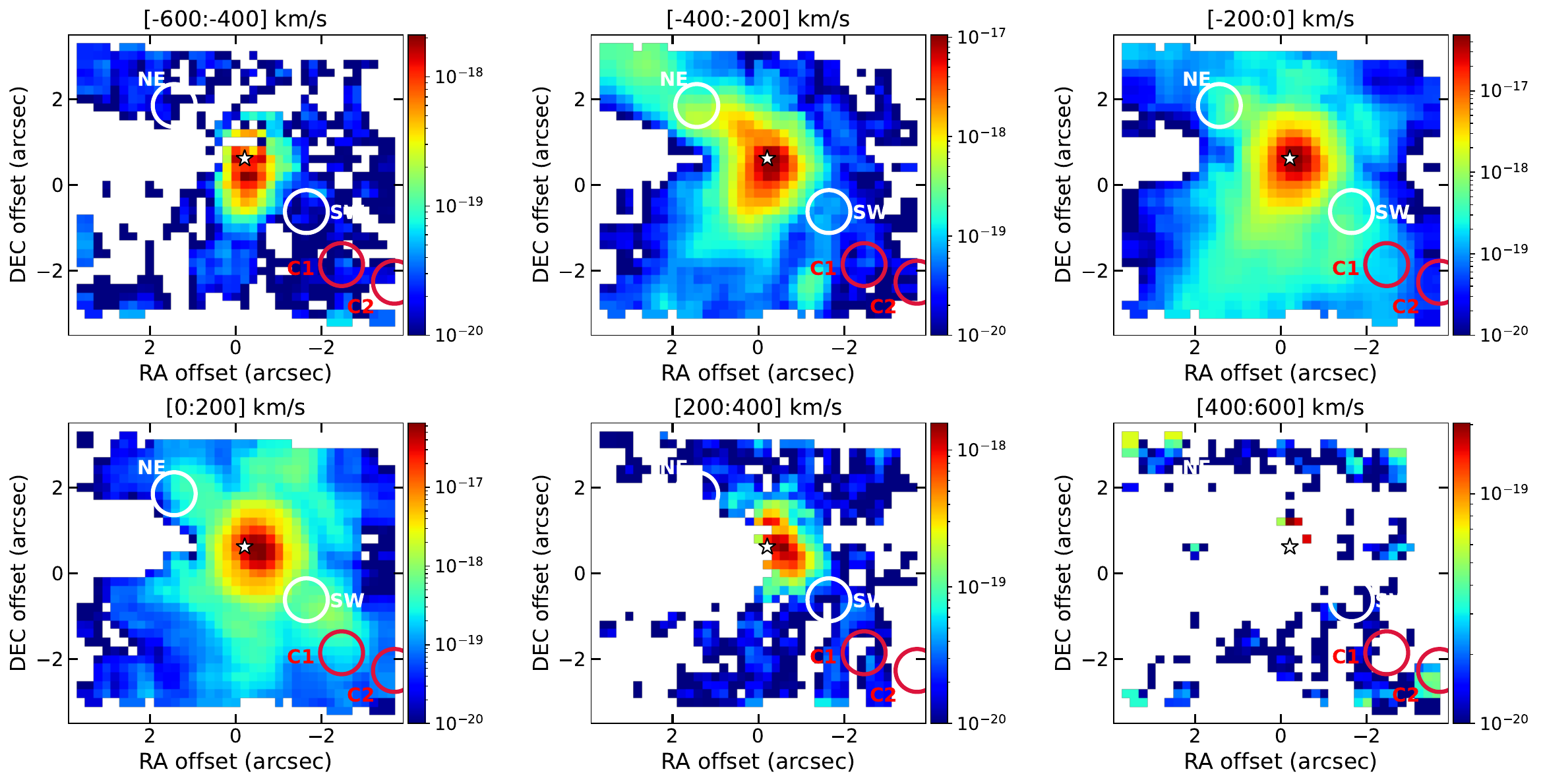}
     \caption{Same as Fig.~\ref{FigAp:ChannelMaps} but for [Ne\,V]$_{14}$ line.}
    \label{FigAp:ChannelMaps_2}
 \end{figure*}

In this appendix we show the continuum maps of channel 3 (see Fig.~\ref{FigAp:ContinuumMaps}), kinematic maps derived using a two component modelling for the emission lines (see Figs.~\ref{FigAp:KinMaps_2comp} and~\ref{FigAp:KinMaps_2comp_2}, and Sect.~\ref{Sect3:Results}), and the channel maps of the neon lines (see Figs.~\ref{FigAp:ChannelMaps},~\ref{FigAp:ChannelMaps_1}, and~\ref{FigAp:ChannelMaps_2}).

When using two Gaussians to fit the [Ne\,V] line (Fig.~\ref{FigAp:KinMaps_2comp_2}), the secondary component is distributed in two regions, one at $\sim 1.5$\arcsec\,NE from the centre, and the second expanding to the SW direction up to $\sim 2.5$\arcsec. The velocity dispersion is similar to that of the primary ($\sigma \leq$300\,km\,s$^{-1}$, see Sect.~\ref{SubSect3:Results_HighIonLines}), which is relatively low, as also found for outflows in other AGNs \citep[see e.g.][]{HM2024}. The region to the NE of the secondary component is completely blueshifted (average velocity $\sim$-200\,km\,s$^{-1}$), whereas the region to the south is both blueshifted and redshifted to the south and west, respectively. We note that these regions are detected with similar properties as the secondary component modelled for the [Ne\,VI], [O\,IV]\,25.89$\mu$m (not shown here), and [S\,IV] emission lines (see Garc{\'i}a-Bernete et al. submitted). As mentioned in Sect.~\ref{SubSect3:Results_HighIonLines}, we detected approaching and receding velocities everywhere in the bi-cone, as is seen from the channel maps in Fig.~\ref{FigAp:ChannelMaps_2}. Thus it is likely that the two components are a consequence of the cone being optically thin in these lines, and we are detecting blue and red wings in the data cubes due to the inclination of the cone with respect to the plane of the sky and its wide angle \citep[see also][]{Roy2021}. 

\begin{figure*}
 \centering
    \includegraphics[width=\textwidth]{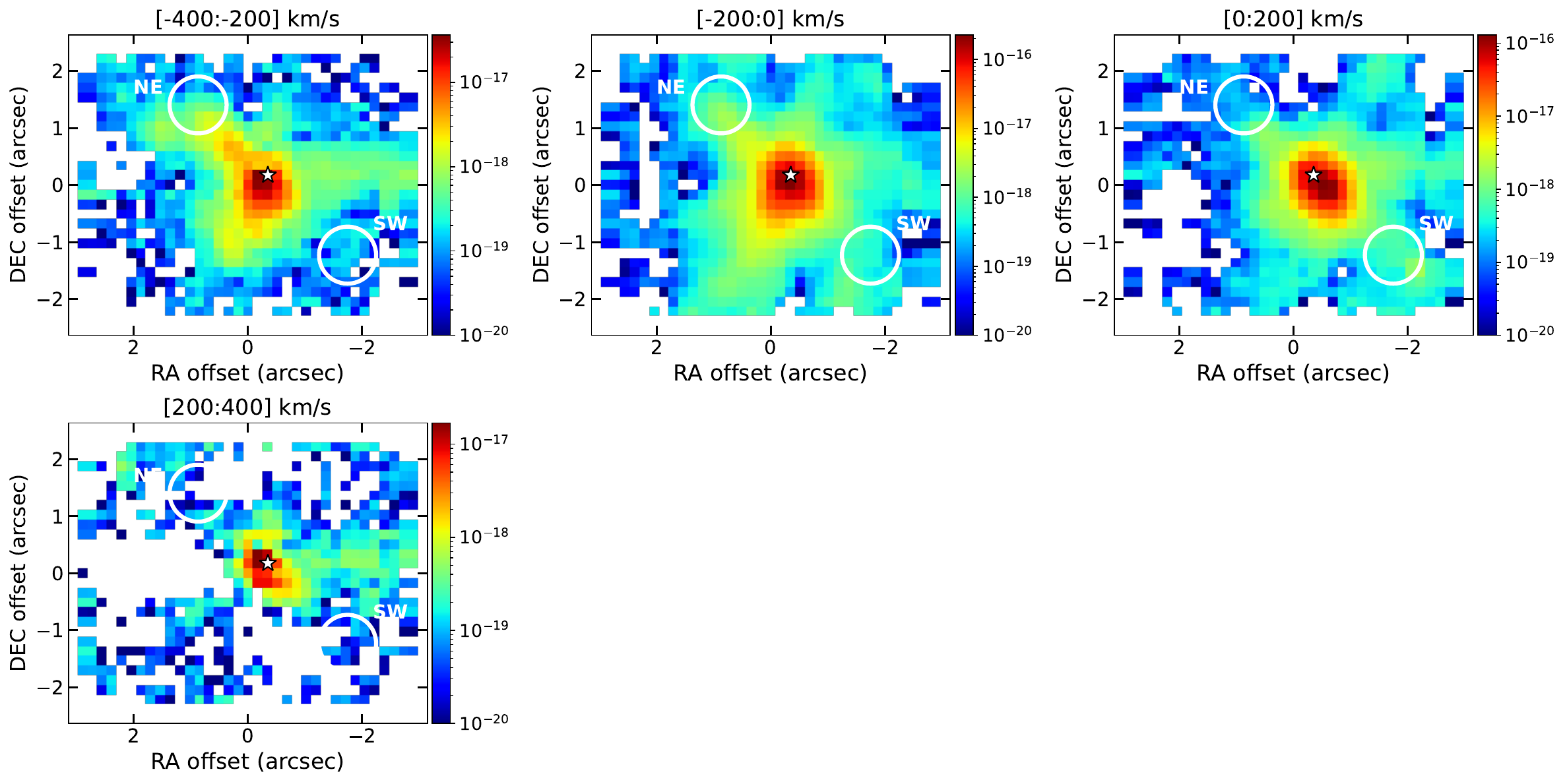}
     \caption{Same as Fig.~\ref{FigAp:ChannelMaps} but for [Ne\,VI] line. We excluded velocities above/below $\pm$\,400\,km\,s$^{-1}$ to eliminate part of the PAH emission, seen as emission in the direction of the disc, where they are more prominent (see Garc{\'i}a-Bernete et al. submitted). }
    \label{FigAp:ChannelMaps_3}
 \end{figure*}

\end{appendix}

\end{document}